\documentclass[reprint, aps, prx, superscriptaddress, twocolumn, showkeys, nobibnotes]{revtex4-2}

\usepackage{graphicx}
\usepackage{dcolumn}
\usepackage{amsmath}
\usepackage{bm}
\usepackage[utf8]{inputenc}
\usepackage{mathtools}
\usepackage{pgffor}
\usepackage[normalem]{ulem}
\usepackage{xcolor}
\usepackage{BOONDOX-cal}
\usepackage{braket}
\usepackage{etaremune}
\usepackage[mediumspace,mediumqspace,squaren]{SIunits}
\usepackage{upgreek}
\usepackage{mathtools}
\usepackage{verbatim}
\usepackage{enumitem}

\usepackage[colorlinks]{hyperref}

\definecolor{mygreen}{RGB}{0, 88, 39}
\definecolor{myblue}{RGB}{46, 48, 146}

\hypersetup{
    colorlinks=true,
    linkcolor=blue,
    filecolor=magenta,      
    urlcolor=blue,
    citecolor=blue
}

\begin{document}

\title{Observation of even-denominator fractional quantum Hall states at $\nu = 3/4$ and 5/4 in the lowest Landau level}
\date{\today}
    
\author{Siddharth Kumar Singh} 
\affiliation{Department of Physics, Columbia University, New York, New York 10027, USA}
\author{Chengyu Wang} 
\affiliation{Department of Electrical and Computer Engineering, Princeton University, Princeton, New Jersey 08544, USA}
\author{Adbhut Gupta}
\affiliation{Department of Electrical and Computer Engineering, Princeton University, Princeton, New Jersey 08544, USA}
\author{Kirk W. Baldwin}
\affiliation{Department of Electrical and Computer Engineering, Princeton University, Princeton, New Jersey 08544, USA}
\author{Loren N. Pfeiffer}
\affiliation{Department of Electrical and Computer Engineering, Princeton University, Princeton, New Jersey 08544, USA}
\author{Mansour Shayegan}
\affiliation{Department of Electrical and Computer Engineering, Princeton University, Princeton, New Jersey 08544, USA}

\let\oldmaketitle\maketitle
\renewcommand\maketitle{{\bfseries\boldmath\oldmaketitle}}
    

\begin{abstract}{Two-dimensional electron systems (2DESs) confined to wide GaAs quantum wells provide a unique platform to study exotic fractional quantum Hall states (FQHSs) because the 2DES has a bilayer charge distribution with significant interlayer tunneling. Precise control over the 2DES density allows the tuning of the interlayer tunneling over a wide range. Here, we present our discovery of new even-denominator FQHSs in the lowest Landau level (orbital index \textit{N} = 0) at filling factors $\nu = 3/4$ and 5/4 in an ultrahigh-quality 2DES confined to a 72.5-nm-wide GaAs quantum well. The ground states at $\nu = 3/4$ and 5/4 both evolve from composite fermion Fermi seas to FQHSs as the density is raised so that interlayer tunneling is sufficiently reduced and the 2DES becomes two-component, signaled by the behavior of the FQHSs flanking $\nu = 3/4$ and 5/4. The two-component nature of the $\nu=3/4$ and 5/4 FQHSs is also evident from their extreme sensitivity to the bilayer charge distribution symmetry: both states disappear quickly when the charge distribution is made asymmetric by only $\simeq 2\%$. We find a natural explanation for the 3/4 and 5/4 FQHSs in terms of two states linked by particle-hole symmetry, and using the Scarola-Jain bilayer composite fermion framework which is a generalization of the well-known, two-component, Halperin state ($\Psi_{331}$ state). Our observations elucidate the crucial role of competing energy and length scales in wide quantum wells in stabilizing new ground states.}
\end{abstract}

\maketitle

\section{Introduction}

Fractional quantum Hall states (FQHSs) are many-body states emerging from the electrons' Coulomb interaction in two-dimensional electron systems (2DESs) subjected to strong perpendicular magnetic fields. FQHSs are typically observed at \textit{odd-denominator} Landau level (LL) filling factors $(\nu)$ in the lowest orbital index (\textit{N} = 0) LL, and can be elegantly understood as the integer quantum Hall states of non-interacting, electron-flux quasiparticles, the composite fermions (CFs) \cite{Jain.PRL.1989, Jain.composite.fermions.2007}. At even-denominator fillings such as $\nu = 1/2, 1/4, ...,$, a Fermi sea is expected and observed \cite{Halperin.Lee.Read.PRB.1993, Willett.PRL.1993}. However, in the excited (\textit{N} = 1) LLs, the residual interaction between CFs stemming from the node in the wavefunction \cite{Willett.PRL.1987, Pan.PRL.1999, Xia.PRL.2004, Scarola.Nature.2000} can result in their pairing instability which manifests itself as \textit{even-denominator} FQHSs. The most notable of these is the FQHS observed at $\nu=5/2$ in GaAs 2DESs \cite{Willett.PRL.1987}. More recently, even-denominator FQHSs were observed in the excited LLs of very high quality samples of ZnO \cite{Falson.Nat.Phys.2015}, monolayer \cite{Kim.NatPhys.2019} and bilayer graphene \cite{Ki.NanoLett.2014, Zibrov.Nature.2017, Li.Science.2017, Huang.PRX.2022, Assouline.PRL.2024, Hu.Nat.Phys.2025}, and monolayer WSe$_2$ \cite{Shi.NatureNanotech.2020}. The even-denominator FQHS in the $N = 1$ LL is generally believed to be a one-component (1C) Pfaffian (or anti-Pfaffian) state with non-Abelian quasi-particles, and have potential use in topological quantum computing \cite{Nayak.RevModPhys.2008}. On the other hand, even-denominator FQHSs in even higher LLs $(N>1)$ \cite{Kim.NatPhys.2019, Hu.Nat.Phys.2025} can verify a new class of wavefunctions that can be understood under the parton paradigm \cite{Jain.PRB.1989, Kim.NatPhys.2019}. 

\begin{figure}[b!]
\hypertarget{fig0}{}
\includegraphics[width=1\columnwidth]{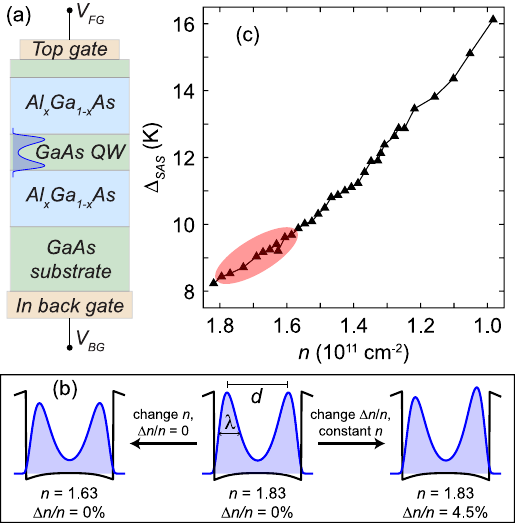}
\caption{(a) Device schematic. (b) Self-consistent Schr\"odinger and Poisson (Hartree) calculations for the charge distribution and potential in a 72.5-nm-wide GaAs QW, showing the effect of changing the density (\textit{n}) and the charge distribution asymmetry ($\Delta n/n$) using the front and back gates. (c) Measured $\Delta_{SAS}$ in our sample plotted as a function of \textit{n}. The shaded, red region marks the range relevant for the data presented in this manuscript.}
\end{figure}

\begin{figure*}[t!]
\hypertarget{fig1}{}
\includegraphics[width=2\columnwidth]{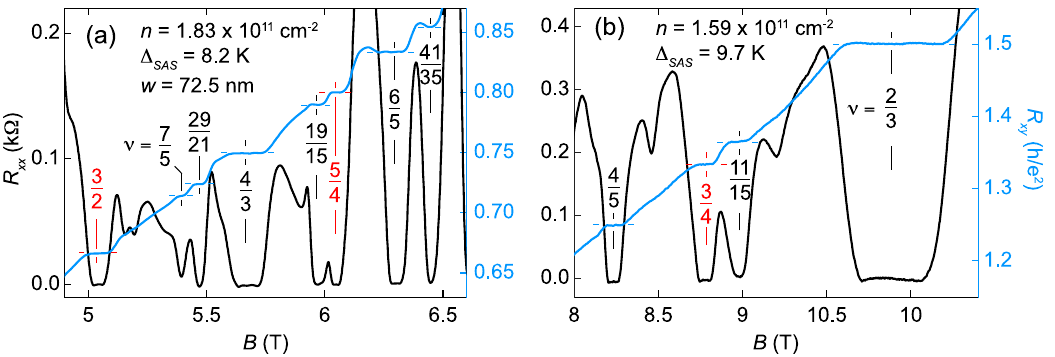}
\caption{(a,b) $R_{xx}$ (black) and $R_{xy}$ (blue) traces taken at $T \simeq 33$ mK, for a 72.5-nm-wide GaAs sample with $n = 1.83$ and 1.59, respectively. Strong even-denominator FQHSs are observed at $\nu = 3/2, 5/4$, and 3/4. Spontaneously imbalanced, bilayer FQHSs are observed in close vicinity of the 5/4 and 3/4 FQHSs at $\nu = 19/15$ and 11/15, respectively. Vertical lines mark the expected positions of the odd- and even-denominator FQHSs, and horizontal lines mark the expected positions for the Hall plateaus.}
\end{figure*}

{Observations of even-denominator FQHSs are therefore always interesting as they offer an alternate route to realizing highly-interacting, many-body states that go beyond Laughlin's wavefunction \cite{Laughlin.PRL.1983}, or the simplest CF phenomenology \cite{Jain.composite.fermions.2007}. Even-denominator FQHSs are also observed in the \textit{N} = 0 LL. The realization of even-denominator FQHSs in the \textit{N} = 0 LL is still understood to rely on CF pairing through the softening of Coulomb interaction of electrons which is achieved via different mechanisms. In systems where the electrons have a bilayer charge distribution, even-denominator FQHSs are observed at $\nu = 1/2$ \cite{Suen.PRL.1992, Eisenstein.PRL.1992, Eisenstein.Annu.Review.2014, Suen2.PRL.1992, Suen.PRL.1994, Liu.PRL.2014, Li.Nat.Phys.2019, Shafayat.PRL.2023, Zhang.Nature.2025} and 1/4 \cite{Luhman.PRL.2008, Shabani.PRL.2009, Shabani2.PRL.2009}. The $\nu = 1/2$ FQHS observed in wide GaAs quantum wells (QWs), at small densities, is expected to be a 1C, Moore-Read Pfaffian-like state mediated by the large interlayer tunneling and electron layer thickness \cite{Park.PRB.1998, Shabani.PRB.2013, Peterson.PRB.2010, Zhu.PRB.2016, Sharma.PRB.2024}; this was verified through mesoscopic experiments that demonstrated 1C Fermi sea of CFs \cite{Mueed.PRL.2015, Mueed.PRL.2016} and the observation of the daughter states of the 1C, Moore-Read Pfaffian state \cite{Singh.NatPhys.2024}. In double-QW GaAs \cite{Eisenstein.PRL.1992, Eisenstein.Annu.Review.2014}, double-layer graphene \cite{Li.Nat.Phys.2019, Zhang.Nature.2025}, and high density electron systems in wide GaAs QWs \cite{Singh.PRL.2025}, the 2C, Halperin-331 state \cite{Halperin.Helv.Phys.Acta.1983} is stabilized. The negligibly small interlayer tunneling in these systems allows the individual layers to act as a pseudospin. Indeed, in multicomponent systems such as spin-coupled subbands \cite{Liu.PRB.2014} or valleys \cite{Zibrov.Nat.Phys.2018}, the 2C, even-denominator FQHS is observed. Yet another mechanism that can also soften the residual interaction between CFs and result in pairing instabilities that manifest even-denominator FQHSs is severe LL mixing \cite{Wang.PRL.2022, Wang.PRL.2023, Wang.PNAS.2023,
Zhao.PRL.2023, Kumar.Nat.Comm.2025, Chen.Nat.Comm.2024, Chanda.Preprint.2025}.}

\begin{figure}[b!]
\hypertarget{fig1_arrhenius}{}
\includegraphics[width=0.8\columnwidth]{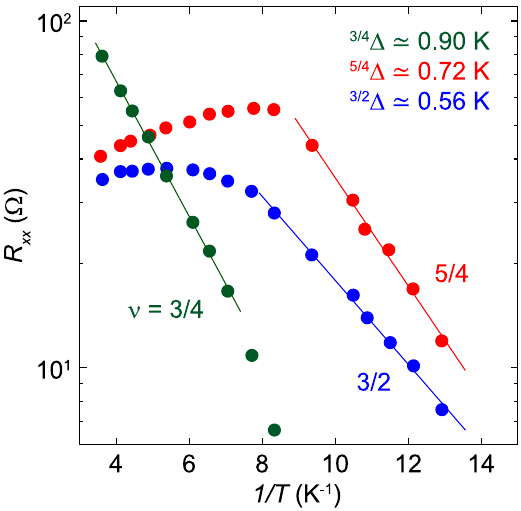}
\caption{Arrhenius plots of the $R_{xx}$ minima at $\nu = 3/2$ (blue), 5/4 (red) and 3/4 (green) FQHSs at $n = 1.83$, allowing us to extract their energy gaps as indicated in the upper right corner.}
\end{figure}

\section{Experimental system}

In wide GaAs QWs, the 2DES has a bilayer charge distribution with tunable interlayer tunneling and layer thickness \cite{Suen.PRL.1994, Shabani.PRB.2013, Singh.NatPhys.2024, Singh.PRL.2025, Manoharan.PRL.1996, Suen.PRB.1991, Shayegan.Review.LesHouches.1999}, rendering this electron system a unique platform for studies of many-body phenomena. Figure \hyperlink{fig0}{1(a)} shows a schematic of our sample which has a front gate as well as a back gate. We control the 2DES density (\textit{n}) as well as the symmetry of the charge distribution $(\Delta n/n)$ in the QW by appropriately biasing the two gates. These operations are demonstrated in Fig. \hyperlink{fig0}{1(b)} using the results of a self-consistent Schr\"odinger and Poisson (Hartree) calculation of the charge distribution and potential in our 72.5-nm-wide GaAs QW. {The center and left panels are for $n = 1.83$ and 1.63 (in units of $10^{11}\ \rm{cm^{-2}}$ which we use throughout this manuscript) when the charge distribution is symmetric i.e., $\Delta n/n = 0$. The right panel shows the results for $n = 1.83$ when $\Delta n/n \simeq 4.5\%$; this is achieved in our device by increasing $n$ by $\Delta n/2$ using one gate and reducing it by $\Delta n/2$ using the other gate, resulting in a total difference of $\Delta n$ charge between the two interfaces of the QW. The interlayer tunneling is quantified by $\Delta_{SAS}$, the difference in the energy levels of the two lowest (symmetric and antisymmetric) electric subbands in the QW when the charge distribution is symmetric.} $\Delta_{SAS}$ can be experimentally extracted from the Fourier transform of the Shubnikov-de Haas oscillations at low magnetic fields, and can be tuned \textit{in-situ} by changing $n$ \cite{Suen.PRB.1991, Suen.PRL.1992, Suen2.PRL.1992, Suen.PRL.1994, Manoharan.PRL.1996, Manoharan.PRL.1997, Shayegan.Review.LesHouches.1999, Shabani.PRB.2013, Singh.NatPhys.2024, Singh.PRL.2025}. The measured $\Delta_{SAS}$ is shown as a function of \textit{n} in Fig. \hyperlink{fig0}{1(c)} for our sample. The shaded red region is the range of density of interest for this manuscript. The interlayer separation $(d)$ sets the strength of the interlayer Coulomb interaction $(e^2/4\pi\epsilon\epsilon_0 d)$, while the parameter $\lambda$, defined as the full-width-at-half-maximum of the individual layers, is vital in modifying (softening) the intralayer Coulomb interaction; $\epsilon\simeq13$ is the GaAs dielectric constant. The parameters $d$ and $\lambda$ are denoted in the center panel of Fig. \hyperlink{fig0}{1(b)}.  

\section{Results}

\begin{figure}[t!]
\hypertarget{fig3}{}
\includegraphics[width=1\columnwidth]{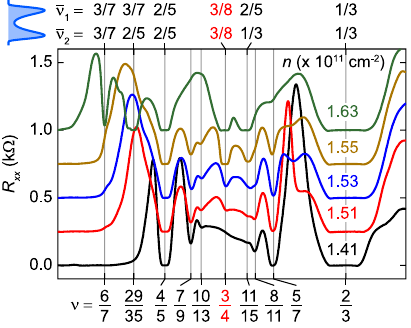}
\caption{Evolution of the ground state at $\nu = 3/4$ from a CF Fermi sea to a FQHS as the density is raised. $R_{xx}$ traces are shifted vertically for clarity. The 2DES becomes 2C at high density, signaled by the disappearance of the 1C, \textit{odd-numerator} FQHSs at $\nu = 5/7$ and 7/9, and the emergence of 2C, bilayer FQHSs at $\nu = 6/7$, 29/35, and 11/15. The layer fillings $\bar{\nu}_1\ \rm{and}\ \bar{\nu}_2$ associated with the observed FQHSs are shown above the figure.}
\end{figure}

\begin{figure}[t!]
\hypertarget{fig4}{}
\includegraphics[width=1\columnwidth]{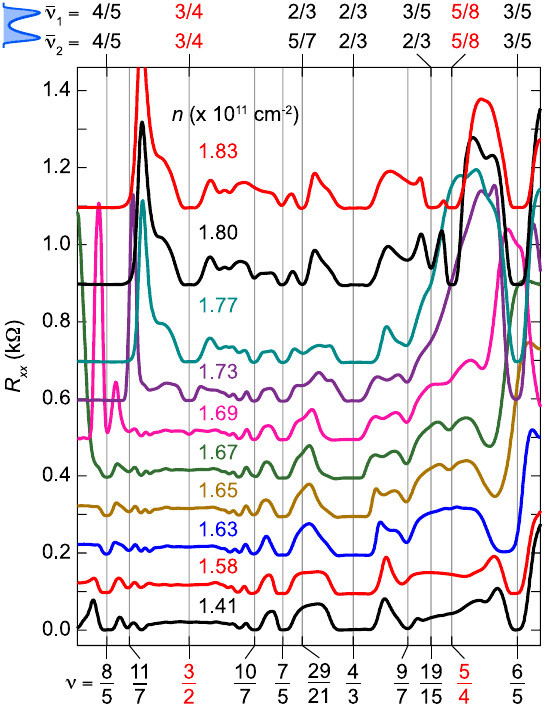}
\caption{Evolution of the 2DES around $\nu = 3/2$ and 5/4 as the density is raised. For clarity, $R_{xx}$ traces are shifted vertically. With increasing \textit{n}, the states at $\nu = 3/2$ and 5/4 evolve from compressible Fermi seas to incompressible FQHSs. 1C to 2C transitions can be observed at the even-numerator $\nu = 4/3$ FQHS at $n = 1.73$, and the $\nu = 6/5$ FQHS at $n = 1.69$. The odd-numerator FQHSs at $\nu = 11/7$, 7/5 and 9/7 become weak or disappear, while new FQHSs emerge at $\nu = 19/15$ and 29/21 at $n = 1.83$.}
\end{figure}

Figure \hyperlink{fig1}{2} highlights our main findings of strong FQHSs at even-denominator filling factors $\nu = 3/2$, 5/4 and 3/4 in an ultrahigh-quality 2DES, with mobility exceeding $1\times10^7$ cm$^{2}$/Vs at $T = 0.3$ K, confined to a 72.5-nm-wide GaAs QW at $n = 1.83$ [Fig. \hyperlink{fig1}{2(a)}] and 1.59 [Fig. \hyperlink{fig1}{2(b)}]. The traces presented in this manuscript are all taken at $T \simeq 33$ mK, unless otherwise stated. All three FQHSs are well developed, as evinced by their vanishing longitudinal resistance $(R_{xx})$ and quantized plateaus in the Hall resistance $(R_{xy})$; the horizontal red lines mark the expected values of quantization for $R_{xy}$ at $h/\nu e^2$. The $R_{xx}$ and $R_{xy}$ traces also show robust FQHSs at even-numerator $\nu = 4/3$ and 6/5 [Fig. \hyperlink{fig1}{2(a)}], and at $\nu = 4/5$ and 2/3 [Fig. \hyperlink{fig1}{2(b)}]. In addition, odd-denominator FQHSs are observed at $\nu = 29/21$, 19/15, and 41/35 [Fig. \hyperlink{fig1}{2(a)}], and at $\nu = 11/15$ [Fig. \hyperlink{fig1}{2(b)}]. {These FQHSs are not observed in single-layer 2DESs in narrow QWs, and their emergence is linked to a spontaneously-imbalanced, bilayer charge distribution in the wide QWs \cite{Manoharan.PRL.1997, Bell.Preprint.2025}; see Supplemental Material (SM) Section S1 for full-field $R_{xx}$ and $R_{xy}$ traces in the low, intermediate and high density regimes \cite{SM.2025}.} 

As is characteristic of FQHSs, $R_{xx}$ minima at $\nu = 3/2$, 5/4 and 3/4 FQHSs are lifted with increasing temperature ($T$), allowing us to extract their energy gaps $(^\nu\Delta)$ from Arrhenius plots. The data shown in Fig. \hyperlink{fig1_arrhenius}{3} were obtained at $n = 1.83$. We extract an energy gap of $^{3/4}\Delta \simeq 0.90$ K, $^{5/4}\Delta \simeq 0.72$ K, and $^{3/2}\Delta \simeq 0.56$ K. To account for the different magnetic fields that these FQHSs occur, we normalize the $^\nu\Delta$ with the Coulomb energy, $E_C = e^2/4\pi\epsilon\epsilon_0\ell_B$, where $\ell_B = \sqrt{\hbar/eB}$ is the magnetic length. In the normalized units, $^{3/4}\Delta \simeq 0.0056\ E_C$, $^{5/4}\Delta \simeq 0.0058\ E_C$ and $^{3/2}\Delta \simeq 0.0049\ E_C$. The 3/2 FQHS was previously reported in lower mobility samples ($\simeq 1\times10^6$ cm$^{2}$/Vs) \cite{Suen.PRL.1994}. However, the emergence of the 3/4 and 5/4 FQHSs evidently requires the highest mobility samples ($\simeq 1\times10^7$ cm$^{2}$/Vs) that have become available only recently \cite{Chung.NatMater.2021, Chung.PRB.2022}. It is noteworthy that, despite the much higher quality samples, the 3/2 FQHS has very similar energy gap to the 3/4 and 5/4 FQHSs, and is in fact marginally weaker. In the older, more disordered samples studied in Refs. \cite{Suen.PRL.1994, Manoharan.PRL.1996, Manoharan.PRL.1997}, the plateaus at $\nu = 4/3$ and 11/15 FQHSs were broad because of localized quasiparticles over a large range of fillings. In the improved samples, smaller disorder landscape presumably allows us to resolve the quarter-filled 3/4 and 5/4 FQHSs.

{The evolution with $n$ of the FQHSs at fillings near $\nu = 3/4$, 5/4 and 3/2 can further illuminate their origin. Focusing first on $\nu = 3/4$, Fig. \hyperlink{fig3}{4} displays the $R_{xx}$ traces taken for different $n$. At $n = 1.41$, a compressible state is observed at $\nu = 3/4$ and is flanked by odd-denominator FQHSs belonging to the Jain sequences at $\nu = 4/5$, 7/9 and 2/3, 5/7, 8/11. This is consistent with the existence of a CF Fermi sea at $\nu = 3/4$. The $R_{xx}$ trace at $n = 1.41$ is characteristic of single-layer 2DESs in narrow GaAs QWs, suggesting the one-component (1C) nature of these FQHSs. As the density is raised, a minimum in $R_{xx}$ at $\nu = 3/4$ emerges $(n = 1.51,\ 1.53)$, and quickly becomes vanishingly small $(n = 1.55,\ 1.63)$. Concomitant with the emergence of the 3/4 FQHS, odd-denominator FQHSs at $\nu = 5/7$ and 7/9 disappear and new odd-denominator FQHSs are observed at $\nu = 6/7$, 29/35 and 11/15, while the FQHSs at $\nu = 2/3$ and 4/5 remain strong in all the traces in Fig. \hyperlink{fig3}{4}. 

The new, odd-denominator FQHSs, which are not observed in single-layer 2DESs \cite{Huang.NatComm.2024}, can be understood as two-component (2C), bilayer FQHSs where each layer stabilizes a Jain-sequence FQHS \cite{spin.polarization.CFs}. We denote the independent layer fillings by $\bar{\nu}_i$, where $i = 1$, 2 is the layer index, and show these above the top x-axis of Fig. \hyperlink{fig3}{4}. For example, the 6/7 FQHS can be understood as two independent layers each forming a 3/7 FQHS. Interestingly, in wide GaAs QWs, FQHSs with \textit{unequal} layer densities (and layer fillings) can also be stabilized by an interaction-induced, spontaneous charge distribution asymmetry \cite{Manoharan.PRL.1997}, such as at $\nu = 29/35$ and 11/15. At $\nu = 11/15$, for e.g., as demonstrated in Ref. \cite{Manoharan.PRL.1997}, the capacitive energy cost of a small, spontaneous, interlayer charge transfer $(\Delta n/n = 1/11)$ can be offset by the condensation of the bilayer system into FQHSs at layer fillings $\bar{\nu}_1 = 1/3$ and $\bar{\nu}_2 = 2/5$. It should be noted that this capacitive energy cost is further reduced because of the large interlayer tunneling in the 2DES realized in wide QWs. The 2C, bilayer nature of the 2DES is further corroborated by the disappearance of the odd-numerator FQHSs at $\nu = 5/7$ and 7/9 as the individual layers are respectively at $\bar{\nu}_{1,2} = 5/14$ and 7/18, where no FQHS is observed in single-layer 2DESs. The even-numerator FQHS at $\nu = 2/3$ and 4/5 can have a 1C or 2C origin \cite{Suen.PRL.1994, Manoharan.PRL.1996, Lay.PRB.1997}, and are strong in Fig. \hyperlink{fig3}{4} traces.

The $R_{xx}$ traces around $\nu = 3/2$ and 5/4, displayed in Fig. \hyperlink{fig4}{5}, follow a qualitatively similar narrative, albeit in a higher density range. Shown on the top x-axis are the individual layer fillings $(\bar{\nu}_{1,2})$ of the prominent FQHSs observed in Fig. \hyperlink{fig4}{5}. At small $n = 1.41$, a compressible Fermi sea of CFs at $\nu = 3/2$ flanked by odd-denominator, Jain sequence FQHSs at $\nu = 4/3$, 7/5, 10/7, and 8/5, 11/7 are observed. At this \textit{n}, we also observe a compressible CF Fermi sea at $\nu = 5/4$ flanked by odd-denominator FQHSs at $\nu = 9/7$ and 6/5. With increasing \textit{n}, the \textit{even-numerator} 4/3 FQHS shows a characteristic 1C to 2C transition in that it is at its weakest around $n = 1.73$; similarly, the 6/5 FQHS disappears at $n = 1.69$ and remerges at $n = 1.73$ signaling its transition from a 1C to a 2C FQHS (see SM Figs. S4, S5 for a phase diagram of the 2D electrons as a function of \textit{n} \cite{SM.2025}). On the other hand, the \textit{odd-numerator} $\nu = 11/7$, 7/9 and 9/7 FQHSs disappear or become weak at the highest density, $n = 1.83$. Also observed is the 19/15 FQHS which is a spontaneously-imbalanced, 2C bilayer FQHS with layer fillings $\bar{\nu}_1 = 2/3$ and $\bar{\nu}_2 = 3/5$ \cite{Manoharan.PRL.1997}. The 11/15 and 19/15 FQHSs are related to each other by particle-hole symmetry, indicating the presence of a discrete layer degree of freedom of the 2DES. Two other spontaneously-imbalanced, 2C, bilayer FQHSs are observed at $n = 1.83$ at $\nu = 29/21\ (\bar{\nu}_1 = 2/3\ \rm{and}\ \bar{\nu}_2 = 5/7)$ in Fig. \hyperlink{fig4}{5}, and at 41/35 $(\bar{\nu}_1 = 2/3\ \rm{and}\ \bar{\nu}_2 = 3/5)$ in Fig. \hyperlink{fig1}{2}. 

The data discussed in Figs. \hyperlink{fig1}{2}-\hyperlink{fig4}{5} show that at small \textit{n}, the 2DES stabilizes 1C FQHSs indicative of a single-layer interaction landscape of the 2DES. At large \textit{n}, 2C, bilayer FQHSs emerge at odd-denominator fillings. These FQHSs are a mixture of 2C, bilayer FQHSs with equal layer fillings ($\nu = 6/5$, 6/7, 4/5) and unequal layer fillings ($\nu = 19/15$, 11/15, 41/35, 29/35, 29/21) suggesting that the 2DES behaves as two, independent electron layers. A 1C to 2C transition of the 2DES for $\nu < 2$ is therefore inferred, as \textit{n} is raised. The monotonic reduction of $\Delta_{SAS}$ as \textit{n} is increased [\hyperlink{fig0}{1(c)}] incentivizes the 1C to 2C transition. The 2DES ground state at different $\nu$ is governed by the interlayer and intralayer Coulomb interactions, and $\Delta_{SAS}$, leading to the 1C to 2C transitions to occur at different $\nu$ for different \textit{n} \cite{Singh.PRL.2025}; also see SM Sections S2 and S3 \cite{SM.2025}.

Given that, at high densities, the 2DES behaves as two independent, non-interacting layers at certain odd-denominator fillings, the observations of the even-denominator FQHSs at $\nu = 3/2$, 5/4 and 3/4 in the 2C regime are surprising. This is because the individual layers would be at $\bar{\nu}_{1,2} = 3/4$, 5/8 and 3/8, respectively, none of which are fillings at which strong FQHSs are observed in single-layer 2DESs \cite{Sajoto.PRB.1990, Pan.PRL.2003, Samkharadze.PRB.2015}. In contrast, the 3/2, 5/4 and 3/4 FQHSs we observe are quite robust and have relatively large energy gaps. We therefore invoke candidate 2C, bilayer FQHSs stabilized by both \textit{interlayer} and \textit{intralayer} correlations, similar to the 2C, Halperin $(\Psi_{331})$ state \cite{Halperin.Helv.Phys.Acta.1983}:

\begin{equation}\label{eqn0}
\begin{split}
    \Psi_{331} =  \prod_{j<k=1}^{N}(z_j - z_k)^3 \prod_{r<s=1}^{N}(w_r - w_s)^3 \prod_{j, r}(z_j - w_r)^1,
\end{split}
\end{equation}
where $z_j \rm{and}\ w_r$ represent the complex coordinates of the electrons in the two layers. For brevity, we are omitting the Gaussian decay part of the wavefunction. Note that the exponents of 3 in the first two terms reflect the 1/3-like intralayer correlations, while the exponent of 1 in the third term indicates that there are also interlayer correlations. This wavefunction describes a 2C FQHS at $\nu = 1/2$ $(\bar{\nu}_1=\bar{\nu}_2=1/4)$ \cite{Halperin.Helv.Phys.Acta.1983}. In the next section, we discuss a generalized framework of the $\Psi_{331}$ wavefunction, introduced by Scarola and Jain \cite{Scarola.PRB.2001}, that can be extended to filling factors other than 1/2. 

\section{Theoretical discussion}

Scarola and Jain \cite{Scarola.PRB.2001} studied the theoretical phase diagram of 2C FQHSs. In this section we introduce the theoretical concepts that are relevant to our experimental results. We adopt the notation $(\nu_1, \nu_2\ |m)$ of Ref. \cite{Scarola.PRB.2001} to refer to 2C, bilayer FQHSs. The state  $(\nu_1, \nu_2\ |m)$ has the following wavefunction \cite{Scarola.PRB.2001}:
\begin{equation}\label{eqn1}
    \Psi_{(\nu_1, \nu_2\ |m)} =  \psi_{\nu_1}[\{z_k\}] \psi_{\nu_2}[\{w_s\}] \prod_{r, j}(z_j - w_r)^m,
\end{equation}
where $\nu_{1, 2}$ account for the intralayer correlations, and \textit{m} accounts for the interlayer correlations. The wavefunctions $\psi_{\nu_i}$ describe primary Jain FQHSs at filling factors $\nu_i = p/(2p+1)$, and $z_i, w_j$ are the positions of the electrons in layers 1, 2, respectively. The wavefunctions $\psi_{\nu_i}$ are given by \cite{Scarola.PRB.2001}:
\begin{equation}\label{eqn2}
    \psi_{p/(2p+1)} = P_{LLL}\Phi_p\prod_{r<s}\left(z_r - z_s\right)^{2}.
\end{equation}
The operator $P_{LLL}$ is the projection operator which projects the wavefunction of $p$ filled LLs of electrons $\Phi_p$ to the lowest LL, while the Jastrow factor $\left(z_r - z_s\right)^{2}$ attaches 2 vortices to each electron to convert them into 2-flux CFs. Note that in this notation, the 2C $\Psi_{331}$ state \cite{Halperin.Helv.Phys.Acta.1983} is the $(1/3, 1/3\ |1)$ state. It should also be emphasized that the parameters $\nu_{1,2}$ and $m$ are related to the total filling factor $\nu$ by the expressions:
\begin{equation}
    \nu = \bar{\nu}_1 + \bar{\nu}_2 = \dfrac{{\nu_2}^{-1} - m}{{\nu_1}^{-1}{\nu_2}^{-1}-m^2} + \dfrac{{\nu_1}^{-1} - m}{{\nu_1}^{-1}{\nu_2}^{-1}-m^2},
\end{equation}
where $\bar{\nu}_{1,2}$ are the individual layer fillings. In the trivial case without any interlayer correlations $(m = 0)$, the individual layer filling $\bar{\nu}_{1,2}$ = $\nu_{1,2}$. 

Table \hyperlink{tab2}{I} summarizes some examples of 2C, bilayer FQHSs with different flavors which can be stabilized at particular $\nu < 1$ that are relevant to our data. These include the following three classes:  (i) balanced, bilayer FQHSs without interlayer correlations (black, $\nu_1 = \nu_2,\ m = 0$); (ii)  imbalanced bilayer FQHSs without interlayer correlations (blue, $\nu_1 \neq \nu_2,\ m = 0$); and (iii) balanced, bilayer FQHSs with interlayer correlation (green and red, $\nu_1 = \nu_2,\ m = 1$) \cite{Halperin.Helv.Phys.Acta.1983, Scarola.PRB.2001}. Our data provide evidence that all three classes of 2C, bilayer FQHSs can be stabilized in our sample. 

\begin{table}[b!]
\hypertarget{tab2}{}
\centering
    \begin{tabular}{wc{0.14\columnwidth} wc{0.14\columnwidth} wc{0.14\columnwidth} wc{0.14\columnwidth} wc{0.14\columnwidth} wc{0.14\columnwidth}}
        \hline\hline
         \boldsymbol{$\nu$} & \boldsymbol{$\nu_1$} & \boldsymbol{$\nu_2$} & \boldsymbol{$m$} &  \boldsymbol{$\bar{\nu}_1$} & \boldsymbol{$\bar{\nu}_2$}\\
         \hline
         6/7 & 3/7 & 3/7 & 0 & 3/7 & 3/7\\
         4/5 & 2/5 & 2/5 & 0 & 2/5 & 2/5\\
         2/3 & 1/3 & 1/3 & 0 & 1/3 & 1/3\\
         \color{blue}29/35 & \color{blue}3/7 & \color{blue}2/5 & \color{blue}0 & \color{blue}3/7 & \color{blue}2/5\\
         \color{blue}11/15 & \color{blue}2/5 & \color{blue}1/3 & \color{blue}0 & \color{blue}2/5 & \color{blue}1/3\\
         \color{mygreen}1/2 & \color{mygreen}1/3 & \color{mygreen}1/3 & \color{mygreen}1 & \color{mygreen}1/4 & \color{mygreen}1/4\\
         \color{red}3/4 & \color{red}3/5 & \color{red}3/5 & \color{red}1 & \color{red}3/8 & \color{red}3/8\\
         \hline\hline
    \end{tabular}
    \caption{Examples of 2C, bilayer FQHSs with different flavors that can be stabilized in wide GaAs QWs. The total filling factor is denoted by $\nu$. $\nu_{1,2}$ denote the intralayer correlations, $m$ represents the interlayer correlation, and $\bar{\nu}_{1,2}$ are the layer fillings which are equal to $\nu_{1,2}$ when $m = 0$. The listed FQHSs are color coded: black and blue represent balanced and imbalanced 2C FQHSs without interlayer correlation ($m = 0$), while green and red are used to denote the even-denominator, 2C FQHSs with interlayer correlation ($m = 1$). The FQHSs where $\bar{\nu}_1 = \bar{\nu}_2$ are destabilized with imposed charge distribution asymmetry, whereas the FQHSs with $\bar{\nu}_1 \neq\ \bar{\nu}_2$ are strengthened.}
\end{table}

To better illustrate the above classes of FQHSs, we use the example of the well-known 2C $\Psi_{331}$ state [Eqn. (\ref{eqn0})]. The $\Psi_{331}$ state, or equivalently $(1/3, 1/3\ |1)$ state in our notation [Eqn. (\ref{eqn1})] refers to a 2C, bilayer FQHS of electrons, and occurs at $\nu = 1/2$ \cite{Halperin.Helv.Phys.Acta.1983}. The intralayer electron correlations are similar to those of the $\nu = 1/3$ FQHS, while $m = 1$ quantifies the interlayer correlations. This state is observed in our 72.5-nm-wide GaAs QW at large $n$ \cite{Singh.PRL.2025}; see SM Sections S2 and S3 for a more in depth discussion \cite{SM.2025}. The 3/2 FQHS [Fig. \hyperlink{fig1}{2(a)}] is likely the particle-hole conjugate of the $(1/3, 1/3\ |1)$ state. The 3/4 and 5/4 FQHSs are also bilayer FQHSs, and are stabilized by interlayer and intralayer correlations \cite{Scarola.PRB.2001}. In this framework, the 3/4 FQHS has a ground state which is described by the $(3/5, 3/5\ |1)$ state, whilst the 5/4 FQHS is the particle-hole conjugate of the $(3/5, 3/5\ |1)$ state \cite{note.2C.4_5.FQHS}. The two $(1/3, 1/3\ |1)$ and $(3/5, 3/5\ |1)$ states are very similar to each other and only differ in the nature of the intralayer correlations of the electrons, which is captured by $\nu_i = 1/3$ and 3/5, respectively.

\section{Effect of charge distribution asymmetry}

\begin{figure*}[t!]
\hypertarget{fig2}{}
\includegraphics[width=2\columnwidth]{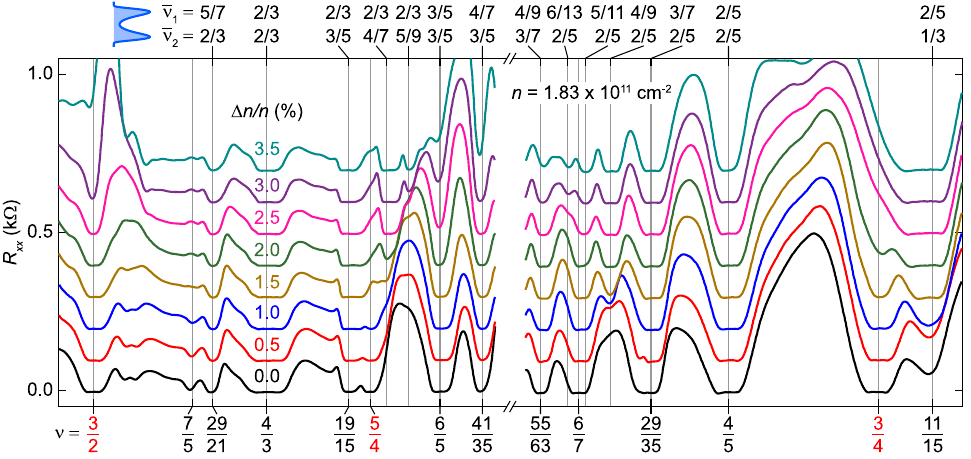}
\caption{$R_{xx}$ traces taken at $n = 1.83$, for different amounts of charge distribution asymmetry $(\Delta n/n)$. Traces are shifted vertically for clarity. Susceptibility of the 3/2, 3/4 and 5/4 FQHSs to $\Delta n/n$ further evinces their 2C, bilayer origin. These FQHSs react to very small changes in $\Delta n/n$, and are destabilized for $\Delta n/n \lesssim 3.5\%$. Balanced, 2C, bilayer FQHSs ($\bar{\nu}_1 = \bar{\nu}_2$) such as those at $\nu = 6/5$, 6/7 and 4/5 are weakened also with increasing $\Delta n/n$, while the imbalanced, 2C, bilayer FQHSs ($\bar{\nu}_1 \neq \bar{\nu}_2$), such as those at $\nu = 11/15$ and 19/15 are typically strengthened.}
\end{figure*}

Next, we test the susceptibility of the FQHSs we observe in our 2DES to imposed charge distribution asymmetry. We achieve this by removing $\Delta n/2$ charge from the back side of the QW using the back gate, and adding $\Delta n/2$ charge to the front side using the front gate, ensuring that the total \textit{n} remains unchanged \cite{Suen.PRB.1991, Shayegan.Review.LesHouches.1999, Suen.PRL.1994, Manoharan.PRL.1996, Manoharan.PRL.1997, Shabani.PRL.2009, Shabani2.PRL.2009, Shabani.PRB.2013, Liu.PRL.2014, Singh.PRL.2025}. The data in Fig. \hyperlink{fig2}{6} indicate that the even-denominator FQHSs at $\nu = 3/2$, 5/4 and 3/4 are strongest when the charge distribution is symmetric, and that they are destabilized even for small values of $\Delta n/n$: The critical charge distribution asymmetries $(\Delta n/n)_C$ for destabilizing the 3/2, 5/4 and 3/4 FQHSs are $^{3/2}(\Delta n/n)_C \simeq 3.5\%$, $^{5/4}(\Delta n/n)_C \simeq 1.5\%$ and $^{3/4}(\Delta n/n)_C \simeq 2.5\%$, respectively, at \textit{n} = 1.83. 

At the same time, the data in Fig. \hyperlink{fig2}{6} are also useful to distinguish between other balanced and imbalanced bilayer FQHSs. The \textit{balanced}, bilayer FQHSs are favored when the layer densities are equal; therefore increasing $\Delta n/n$ weakens these FQHSs. For example, at $\nu = 6/5$ the $R_{xx}$ minimum is lifted for $\Delta n/n\geq2\%$, and at $\nu = 4/5$, 6/7 the width in \textit{B}-field for which the $R_{xx}$ is vanishingly small becomes narrower, as $\Delta n/n$ increases. 

The above behavior is in contrast to what is seen for the spontaneously-$imbalanced$, bilayer FQHSs. While these FQHSs are present even for symmetric charge distributions (at zero magnetic field), they are typically strongest when the charge distribution is made asymmetric (via applying back and front gate biases) so that the individual layer fillings match the filling at which a primary Jain FQHS can be stabilized, and then get weak again with increasing imposed asymmetry \cite{Manoharan.PRL.1997}. An example is the evolution of the 11/15 FQHS $(\bar{\nu}_1=1/3,\ \bar{\nu}_2=2/5)$ which should be most stable when a $\Delta n/n\simeq9.1\%$ is imposed. As seen in Fig. \hyperlink{fig2}{6}, the 11/15 FQHS exhibits only a weak $R_{xx}$ minimum for $\Delta n/n = 0$, and increasing $\Delta n/n$ indeed leads to a strengthening of the 11/15 FQHS, reflected by a vanishing and wide resistance minimum. The 41/35 FQHS $(\bar{\nu}_1=3/5,\ \bar{\nu}_2=4/7)$, and the 55/63 FQHS $(\bar{\nu}_1=3/7,\ \bar{\nu}_2=4/9)$, on the other hand, require very small values of imposed asymmetry for their maximum stability, namely $\Delta n/n \simeq 2.4\%$ and $\Delta n/n \simeq 1.8\%$, respectively. Consistent with this expectation, these two FQHSs are at their maximum strength in Fig. \hyperlink{fig2}{6} when $\Delta n/n \simeq 2\%$.

\section{Concluding remarks}

We reiterate that the data presented in Figs. \hyperlink{fig1}{2}-\hyperlink{fig2}{6} provide evidence for the observation of 2C, layer-balanced $(\bar{\nu}_1=\bar{\nu}_2)$ FQHSs at even-denominator fillings $\nu = 3/4$, 5/4, and 3/2 in wide GaAs QWs at sufficiently high densities. We identify these as 2C FQHSs with interlayer correlations, qualitatively similar to the Halperin $\Psi_{331}$ state, which is generally believed to describe the FQHS at $\nu = 1/2$ observed in GaAs double-QW samples \cite{Eisenstein.PRL.1992}, or in wide QW samples with appropriate parameters \cite{Suen.PRL.1994, Singh.PRL.2025}. 

It is worth contrasting the even-denominator FQHSs observed so far in wide GaAs QWs, namely at $\nu = 3/2$, 5/4, 3/4, 1/2, and 1/4. In all cases, a CF Fermi sea is seen at very low densities, evolving to a FQHS as \textit{n} is increased. In the case of $\nu=1/2$, data suggest that the FQHS is a 1C state at intermediate densities while at higher densities it appears to make a transition to a 2C state \cite{Singh.PRL.2025} before being engulfed by an insulating phase, presumably a bilayer Wigner crystal state \cite{Manoharan.PRL.1996, Hatke.PRB.2017}. The case of the $\nu=1/4$ FQHS is less clear. There is very limited experimental data, partly because of the much larger magnetic fields required to reach $\nu = 1/4$ \cite{Luhman.PRL.2008, Shabani.PRL.2009, Shabani2.PRL.2009}, but theory suggests that it is a 1C FQHS \cite{Faugno.PRL.2019, Sharma.PRB.2024}. In contrast to the cases of 1/2 and 1/4, the 3/2 and 3/4 FQHSs appear to make a transition from the Fermi sea directly to a 2C FQHS as the density is raised. The evolution of the 5/4 FQHS, on the other hand, is more subtle as we now describe.

As depicted in Fig. \hyperlink{fig4}{5}, the ground state at $\nu=5/4$ exhibits a more nuanced evolution compared to $\nu = 3/4$. At $\nu=3/4$, the ground state slowly and monotonically evolves from a CF Fermi sea at the lowest \textit{n} to a FQHS at the highest \textit{n} (Fig. \hyperlink{fig3}{4}). However, the evolution at $\nu=5/4$ does not appear to be monotonic. As seen in Fig. \hyperlink{fig4}{5}, at the lowest densities ($n < 1.63$) the ground state at $\nu=5/4$ is a compressible CF Fermi sea. As we raise \textit{n}, in a narrow intermediate density range ($1.65 \leq n \leq 1.69$), we see evidence for a developing 5/4 FQHS in the form of a weak minimum in $R_{xx}$. As shown in SM Section S4, Fig. S8 \cite{SM.2025}, in this density range the derivative of $R_{xy}$ also dips below the classical Hall slope, towards zero, signaling a developing FQHS \cite{Goldman.PRL.1988}. As we further increase \textit{n}, this weak, developing 5/4 FQHS disappears ($n = 1.73, 1.77$), and then emerges at the highest densities $(n = 1.80, 1.83)$ as a robust, 2C, bilayer 5/4 FQHS. Interestingly, the developing 5/4 FQHS at $n \simeq 1.67$ is observed when the 2DES is still 1C, as signaled by the presence of the nearby, odd-numerator FQHS at $\nu = 9/7$ and the absence of the 2C, $\nu=19/15$ FQHS (Fig. \hyperlink{fig4}{5}). Also, the $R_{xx}$ minimum observed at $\nu = 5/4$ for $n \simeq 1.67$ quickly disappears when $T$ is raised, suggesting its many- body origin (see SM Section S4, Fig. S7 \cite{SM.2025}).

We speculate that the evolution at $\nu = 5/4$ in Fig. \hyperlink{fig4}{5} might suggest the existence of \textit{two distinct} types of FQHSs. While, as demonstrated here (e.g., Fig. \hyperlink{fig2}{6}), the robust $\nu=5/4$ FQHS observed at the highest densities is a 2C state, the 1C FQHSs observed near 5/4 at $n \simeq 1.67$ suggest that the fragile 5/4 FQHS itself may have a 1C origin. Such a 1C FQHS is possibly related to the $\nu = 1/4$ FQHS that has been reported in wide GaAs QWs \cite{Luhman.PRL.2008, Shabani.PRL.2009, Shabani2.PRL.2009}, and is theoretically proposed to have a 1C, non-Abelian origin \cite{Faugno.PRL.2019, Sharma.PRB.2024}. While the 1/4 FQHS is observed in the 1/4-filled $N=0$ LL of the \textit{symmetric} electric subband, the 5/4 FQHS would correspond to the 1/4 filling of the $N = 0$ LL of the \textit{antisymmetric} electric subband. Clearly, however, additional experimental data, e.g., in wide QWs with different width and density, are needed to test our speculation. 

There is also a remarkable difference between the evolutions of the FQHSs at $\nu=6/5$ and 4/5 (Figs. \hyperlink{fig3}{4} and \hyperlink{fig4}{5}). One would expect that both of these states should make a transition from a 1C FQHS at low \textit{n} to a 2C FQHS at high \textit{n}, with 3/5 and 2/5 fillings in each layer, respectively. We observe such a transition for the $\nu=6/5$ FQHS, signaled by a weakening of the $R_{xx}$ minimum at 6/5 at intermediate densities (see Fig. \hyperlink{fig4}{5} and also Figs. S4 and S5). In contrast, the $\nu=4/5$ FQHS remains strong in the entire range of \textit{n}, from the 1C regime and deep into the 2C regime (Figs. \hyperlink{fig3}{4}, S4 and S5). This is puzzling. {However, it is worth remarking that non-Abelian, 2C, bilayer FQHSs have been theoretically suggested \cite{Barkeshli.PRB.2010, Barkeshli2.PRB.2010, Peterson.PRB.2015}, including at $\nu = 4/5$.} The unusual evolution of the 4/5 FQHS as a function of \textit{n} that we observe can be indicative of such an exotic origin, but future work is needed to confirm this association. 

Finally, bilayer FQHSs with interlayer correlations have also been observed in double-layer graphene recently \cite{Li.Nat.Phys.2019, Zhang.Nature.2025}, and have been studied numerically \cite{Faugno.PRB.2020}. Interestingly, the data in Refs. \cite{Li.Nat.Phys.2019, Zhang.Nature.2025} display a large sequence of bilayer FQHSs of the form $({\nu}_1, {\nu}_2\ |1)$. The samples studied in Refs. \cite{Li.Nat.Phys.2019, Zhang.Nature.2025} have very small $d/\ell_B \leq 1$, and negligible interlayer tunneling. In contrast, we only observe bilayer FQHSs of the form $({\nu}_1, {\nu}_2\ |1)$ at $\nu = 3/2$, 5/4 and 3/4. This is likely because of the vastly different energy and length scales in the two platforms. In our electron system, there is large interlayer tunneling of the order of 10 K, Fig. \hyperlink{fig0}{1(c)}, with large interlayer separation $(d/\ell_B \simeq 4-6)$ and significant electron layer thickness $(\lambda)$, as depicted in Fig. \hyperlink{fig0}{1(b)}. It has been suggested that large interlayer tunneling can shift the stability of bilayer QHSs to larger values of $d/\ell_B$ \cite{Murphy.PRL.1994, Eisenstein.Annu.Review.2014, Girvin.MacDonald.chapter, Yang.PRB.1996}; however, each filling factor must be treated on a case-by-case basis, as shown in Ref. \cite{Faugno.PRB.2020} (see also SM Section S3 \cite{SM.2025}). These competing energy and length scales manifest rich sets of FQHSs, and in essence, make unexplored regions of the bilayer FQHS phase-space with significant interlayer tunneling accessible.

\begin{acknowledgments}

We acknowledge support by the National Science Foundation (NSF) Grant No. DMR 2104771 for measurements, and the Gordon and Betty Moore Foundation’s EPiQS Initiative (Grant No. GBMF9615.01 to L.N.P.) for sample fabrication. A portion of this work was performed at the National High Magnetic Field Laboratory, which is supported by National Science Foundation Cooperative Agreement No. DMR-2128556, and the State of Florida. This research is funded in part by QuantEmX travel grants from ICAM and the Gordon and Betty Moore Foundation through Grant GBMF9616 to S.K.S., C.W. and A.G. We thank R. Nowell, G. Jones, A. Bangura and T. Murphy at NHMFL for technical assistance, and J. K. Jain and M. Gattu for illuminating discussions. 

\end{acknowledgments}

\bibliography{references_arxiv.bib}

@article{Shabani.PRB.2013,
  title     = "{Phase diagrams for the stability of the $\ensuremath{\nu}=\frac{1}{2}$ fractional quantum Hall effect in electron systems confined to symmetric, wide GaAs quantum wells}",
  author    = {Shabani, J. and Liu, Yang and Shayegan, M. and Pfeiffer, L. N. and West, K. W. and Baldwin, K. W.},
  journal   = {Phys. Rev. B},
  volume    = {88},
  issue     = {24},
  pages     = {245413},
  numpages  = {8},
  year      = {2013},
  month     = {Dec},
  publisher = {American Physical Society},
  doi       = {10.1103/PhysRevB.88.245413},
  url       = {https://link.aps.org/doi/10.1103/PhysRevB.88.245413}
}

@article{Mueed.PRL.2016,
  title     = "{Geometric Resonance of Composite Fermions near Bilayer Quantum Hall States}",
  author    = {Mueed, M. A. and Kamburov, D. and Pfeiffer, L. N. and West, K. W. and Baldwin, K. W. and Shayegan, M.},
  journal   = {Phys. Rev. Lett.},
  volume    = {117},
  issue     = {24},
  pages     = {246801},
  numpages  = {5},
  year      = {2016},
  month     = {Dec},
  publisher = {American Physical Society},
  doi       = {10.1103/PhysRevLett.117.246801},
  url       = {https://link.aps.org/doi/10.1103/PhysRevLett.117.246801}
}

@article{Mueed.PRL.2015,
  title     = "{Geometric Resonance of Composite Fermions Near the $\ensuremath{\nu}=1/2$ Fractional Quantum Hall State}",
  author    = {Mueed, M. A. and Kamburov, D. and Hasdemir, S. and Shayegan, M. and Pfeiffer, L. N. and West, K. W. and Baldwin, K. W.},
  journal   = {Phys. Rev. Lett.},
  volume    = {114},
  issue     = {23},
  pages     = {236406},
  numpages  = {5},
  year      = {2015},
  month     = {Jun},
  publisher = {American Physical Society},
  doi       = {10.1103/PhysRevLett.114.236406},
  url       = {https://link.aps.org/doi/10.1103/PhysRevLett.114.236406}
}

@article{Peterson.PRB.2010,
  title     = "{Quantum Hall phase diagram of half-filled bilayers in the lowest and the second orbital Landau levels: Abelian versus non-Abelian incompressible fractional quantum Hall states}",
  author    = {Peterson, Michael R. and Das Sarma, S.},
  journal   = {Phys. Rev. B},
  volume    = {81},
  issue     = {16},
  pages     = {165304},
  numpages  = {17},
  year      = {2010},
  month     = {Apr},
  publisher = {American Physical Society},
  doi       = {10.1103/PhysRevB.81.165304},
  url       = {https://link.aps.org/doi/10.1103/PhysRevB.81.165304}
}

@article{Zhu.PRB.2016,
  title     = "{Fractional quantum Hall bilayers at half filling: Tunneling-driven non-Abelian phase}",
  author    = {Zhu, W. and Liu, Zhao and Haldane, F. D. M. and Sheng, D. N.},
  journal   = {Phys. Rev. B},
  volume    = {94},
  issue     = {24},
  pages     = {245147},
  numpages  = {16},
  year      = {2016},
  month     = {Dec},
  publisher = {American Physical Society},
  doi       = {10.1103/PhysRevB.94.245147},
  url       = {https://link.aps.org/doi/10.1103/PhysRevB.94.245147}
}

@article{Sharma.PRB.2024,
  title     = "{Composite-fermion pairing at half-filled and quarter-filled lowest Landau level}",
  author    = {Sharma, Anirban and Balram, Ajit C. and Jain, J. K.},
  journal   = {Phys. Rev. B},
  volume    = {109},
  issue     = {3},
  pages     = {035306},
  numpages  = {22},
  year      = {2024},
  month     = {Jan},
  publisher = {American Physical Society},
  doi       = {10.1103/PhysRevB.109.035306},
  url       = {https://link.aps.org/doi/10.1103/PhysRevB.109.035306}
}

@article{Suen.PRL.1992,
  title     = "{Observation of a \ensuremath{\nu} = 1/2 fractional quantum Hall state in a double-layer electron system}",
  author    = {Suen, Y. W. and Engel, L. W. and Santos, M. B. and Shayegan, M. and Tsui, D. C.},
  journal   = {Phys. Rev. Lett.},
  volume    = {68},
  issue     = {9},
  pages     = {1379--1382},
  numpages  = {0},
  year      = {1992},
  month     = {Mar},
  publisher = {American Physical Society},
  doi       = {10.1103/PhysRevLett.68.1379},
  url       = {https://link.aps.org/doi/10.1103/PhysRevLett.68.1379}
}

@article{Suen2.PRL.1992,
  title     = "{Correlated states of an electron system in a wide quantum well}",
  author    = {Suen, Y. W. and Santos, M. B. and Shayegan, M.},
  journal   = {Phys. Rev. Lett.},
  volume    = {69},
  issue     = {24},
  pages     = {3551--3554},
  numpages  = {0},
  year      = {1992},
  month     = {Dec},
  publisher = {American Physical Society},
  doi       = {10.1103/PhysRevLett.69.3551},
  url       = {https://link.aps.org/doi/10.1103/PhysRevLett.69.3551}
}

@article{Suen.PRL.1994,
  title     = "{Origin of the \ensuremath{\nu} = 1/2 fractional quantum Hall state in wide single quantum wells}",
  author    = {Suen, Y. W. and Manoharan, H. C. and Ying, X. and Santos, M. B. and Shayegan, M.},
  journal   = {Phys. Rev. Lett.},
  volume    = {72},
  issue     = {21},
  pages     = {3405--3408},
  numpages  = {0},
  year      = {1994},
  month     = {May},
  publisher = {American Physical Society},
  doi       = {10.1103/PhysRevLett.72.3405},
  url       = {https://link.aps.org/doi/10.1103/PhysRevLett.72.3405}
}

@article{Manoharan.PRL.1996,
  title     = "{Evidence for a Bilayer Quantum Wigner Solid}",
  author    = {Manoharan, H. C. and Suen, Y. W. and Santos, M. B. and Shayegan, M.},
  journal   = {Phys. Rev. Lett.},
  volume    = {77},
  issue     = {9},
  pages     = {1813--1816},
  numpages  = {0},
  year      = {1996},
  month     = {Aug},
  publisher = {American Physical Society},
  doi       = {10.1103/PhysRevLett.77.1813},
  url       = {https://link.aps.org/doi/10.1103/PhysRevLett.77.1813}
}

@article{Hatke.PRB.2017,
  title     = "{Microwave spectroscopic observation of a Wigner solid within the $\ensuremath{\nu}=1/2$ fractional quantum Hall effect}",
  author    = {Hatke, A. T. and Liu, Yang and Engel, L. W. and Pfeiffer, L. N. and West, K. W. and Baldwin, K. W. and Shayegan, M.},
  journal   = {Phys. Rev. B},
  volume    = {95},
  issue     = {4},
  pages     = {045417},
  numpages  = {5},
  year      = {2017},
  month     = {Jan},
  publisher = {American Physical Society},
  doi       = {10.1103/PhysRevB.95.045417},
  url       = {https://link.aps.org/doi/10.1103/PhysRevB.95.045417}
}

@article{Singh.NatPhys.2024,
  title     = "{Topological phase transition between Jain states and daughter states of the $\nu$ = 1/2 fractional quantum Hall state}",
  author    = {Singh, S. K. and Wang, C and Tai, C. T. and Calhoun, C. S. and Villegas Rosales, K. A. and Madathil, P. T. and Gupta, A. and Baldwin, K. W. and Pfeiffer, L. N. and Shayegan, M.},
  journal   = {Nat. Phys.},
  volume    = {20},
  pages     = {1247--1252},
  numpages  = {0},
  year      = {2024},
  month     = {May},
  publisher = {Nature Publishing Group UK London},
  doi       = {10.1038/s41567-024-02517-w},
  url       = {https://www.nature.com/articles/s41567-024-02517-w#citeas}
}

@article{Assouline.PRL.2024,
  title     = "{Energy Gap of the Even-Denominator Fractional Quantum Hall State in Bilayer Graphene}",
  author    = {Assouline, Alexandre and Wang, Taige and Zhou, Haoxin and Cohen, Liam A. and Yang, Fangyuan and Zhang, Ruining and Taniguchi, Takashi and Watanabe, Kenji and Mong, Roger S. K. and Zaletel, Michael P. and Young, Andrea F.},
  journal   = {Phys. Rev. Lett.},
  volume    = {132},
  issue     = {4},
  pages     = {046603},
  numpages  = {6},
  year      = {2024},
  month     = {Jan},
  publisher = {American Physical Society},
  doi       = {10.1103/PhysRevLett.132.046603},
  url       = {https://link.aps.org/doi/10.1103/PhysRevLett.132.046603}
}

@article{Chung.NatMater.2021,
  title     = "{Ultra-high-quality two-dimensional electron systems}",
  author    = {Chung, Yoon Jang and Villegas Rosales, KA and Baldwin, KW and Madathil, PT and West, KW and Shayegan, M and Pfeiffer, LN},
  journal   = {Nat. Mater.},
  volume    = {20},
  number    = {5},
  pages     = {632--637},
  year      = {2021},
  publisher = {Nature Publishing Group UK London},
  doi       = {10.1038/s41563-021-00942-3},
  url       = {https://www.nature.com/articles/s41563-021-00942-3#citeas}
}

@article{Willett.PRL.1987,
  title     = "{Observation of an even-denominator quantum number in the fractional quantum Hall effect}",
  author    = {Willett, R. and Eisenstein, J. P. and St\"ormer, H. L. and Tsui, D. C. and Gossard, A. C. and English, J. H.},
  journal   = {Phys. Rev. Lett.},
  volume    = {59},
  issue     = {15},
  pages     = {1776--1779},
  numpages  = {0},
  year      = {1987},
  month     = {Oct},
  publisher = {American Physical Society},
  doi       = {10.1103/PhysRevLett.59.1776},
  url       = {https://link.aps.org/doi/10.1103/PhysRevLett.59.1776}
}

@article{Scarola.Nature.2000,
  title     = "{Cooper instability of composite fermions}",
  author    = {Scarola, Vito W and Park, Kwon and Jain, Jainendra K},
  journal   = {Nature},
  volume    = {406},
  number    = {6798},
  pages     = {863--865},
  year      = {2000},
  publisher = {Nature Publishing Group UK London},
  doi       = {10.1038/35022524},
  url       = {https://www.nature.com/articles/35022524#citeas}
}

@article{Eisenstein.PRL.1992,
  title     = "{New fractional quantum Hall state in double-layer two-dimensional electron systems}",
  author    = {Eisenstein, J. P. and Boebinger, G. S. and Pfeiffer, L. N. and West, K. W. and He, Song},
  journal   = {Phys. Rev. Lett.},
  volume    = {68},
  issue     = {9},
  pages     = {1383--1386},
  numpages  = {0},
  year      = {1992},
  month     = {Mar},
  publisher = {American Physical Society},
  doi       = {10.1103/PhysRevLett.68.1383},
  url       = {https://link.aps.org/doi/10.1103/PhysRevLett.68.1383}
}

@article{Halperin.Helv.Phys.Acta.1983,
  title     = "{Theory of the quantized Hall conductance}",
  author    = {Halperin, Bertrand I},
  journal   = {Helv. Phys. Acta},
  volume    = {56},
  number    = {1-3},
  pages     = {75},
  year      = {1983}
}

@article{Barkeshli.PRB.2010,
  title     = "{Classification of Abelian and non-Abelian multilayer fractional quantum Hall states through the pattern of zeros}",
  author    = {Barkeshli, Maissam and Wen, Xiao-Gang},
  journal   = {Phys. Rev. B},
  volume    = {82},
  issue     = {24},
  pages     = {245301},
  numpages  = {19},
  year      = {2010},
  month     = {Dec},
  publisher = {American Physical Society},
  doi       = {10.1103/PhysRevB.82.245301},
  url       = {https://link.aps.org/doi/10.1103/PhysRevB.82.245301}
}

@article{Barkeshli2.PRB.2010,
  title     = "{Non-Abelian two-component fractional quantum Hall states}",
  author    = {Barkeshli, Maissam and Wen, Xiao-Gang},
  journal   = {Phys. Rev. B},
  volume    = {82},
  issue     = {23},
  pages     = {233301},
  numpages  = {4},
  year      = {2010},
  month     = {Dec},
  publisher = {American Physical Society},
  doi       = {10.1103/PhysRevB.82.233301},
  url       = {https://link.aps.org/doi/10.1103/PhysRevB.82.233301}
}

@article{Nayak.RevModPhys.2008,
  title     = "{Non-Abelian anyons and topological quantum computation}",
  author    = {Nayak, Chetan and Simon, Steven H. and Stern, Ady and Freedman, Michael and Das Sarma, Sankar},
  journal   = {Rev. Mod. Phys.},
  volume    = {80},
  issue     = {3},
  pages     = {1083--1159},
  numpages  = {0},
  year      = {2008},
  month     = {Sep},
  publisher = {American Physical Society},
  doi       = {10.1103/RevModPhys.80.1083},
  url       = {https://link.aps.org/doi/10.1103/RevModPhys.80.1083}
}

@article{Wang.PRL.2022,
  title     = "{Even-Denominator Fractional Quantum Hall State at Filling Factor $\ensuremath{\nu}=3/4$}",
  author    = {Wang, Chengyu and Gupta, A. and Singh, S. K. and Chung, Y. J. and Pfeiffer, L. N. and West, K. W. and Baldwin, K. W. and Winkler, R. and Shayegan, M.},
  journal   = {Phys. Rev. Lett.},
  volume    = {129},
  issue     = {15},
  pages     = {156801},
  numpages  = {6},
  year      = {2022},
  month     = {Oct},
  publisher = {American Physical Society},
  doi       = {10.1103/PhysRevLett.129.156801},
  url       = {https://link.aps.org/doi/10.1103/PhysRevLett.129.156801}
}

@article{Wang.PRL.2023,
  title     = "{Fractional Quantum Hall State at Filling Factor $\ensuremath{\nu}=1/4$ in Ultra-High-Quality GaAs Two-Dimensional Hole Systems}",
  author    = {Wang, Chengyu and Gupta, A. and Singh, S. K. and Madathil, P. T. and Chung, Y. J. and Pfeiffer, L. N. and Baldwin, K. W. and Winkler, R. and Shayegan, M.},
  journal   = {Phys. Rev. Lett.},
  volume    = {131},
  issue     = {26},
  pages     = {266502},
  numpages  = {7},
  year      = {2023},
  month     = {Dec},
  publisher = {American Physical Society},
  doi       = {10.1103/PhysRevLett.131.266502},
  url       = {https://link.aps.org/doi/10.1103/PhysRevLett.131.266502}
}

@article{Wang.PNAS.2023,
author      = {Chengyu Wang  and Adbhut Gupta  and Pranav T. Madathil  and Siddharth K. Singh  and Yoon Jang Chung  and Loren N. Pfeiffer  and Kirk W. Baldwin  and Mansour Shayegan },
title       = "{Next-generation even-denominator fractional quantum Hall states of interacting composite fermions}",
journal     = {Proc. Natl. Acad. Sci. U.S.A.},
volume      = {120},
number      = {52},
pages       = {e2314212120},
year        = {2023},
doi         = {10.1073/pnas.2314212120},
URL         = {https://www.pnas.org/doi/abs/10.1073/pnas.2314212120}
}

@article{Shi.NatureNanotech.2020,
  title     = "{Odd-and even-denominator fractional quantum Hall states in monolayer WSe2}",
  author    = {Shi, Qianhui and Shih, En-Min and Gustafsson, Martin V and Rhodes, Daniel A and Kim, Bumho and Watanabe, Kenji and Taniguchi, Takashi and Papi{\'c}, Zlatko and Hone, James and Dean, Cory R},
  journal   = {Nat. Nanotechnol.},
  volume    = {15},
  number    = {7},
  pages     = {569--573},
  year      = {2020},
  publisher = {Nature Publishing Group UK London},
  doi       = {10.1038/s41565-020-0685-6},
  url       = {https://www.nature.com/articles/s41565-020-0685-6#citeas}
}

@article{Kim.NatPhys.2019,
  title     = "{Even denominator fractional quantum Hall states in higher Landau levels of graphene}",
  author    = {Kim, Youngwook and Balram, Ajit C and Taniguchi, Takashi and Watanabe, Kenji and Jain, Jainendra K and Smet, Jurgen H},
  journal   = {Nat. Phys.},
  volume    = {15},
  number    = {2},
  pages     = {154--158},
  year      = {2019},
  publisher = {Nature Publishing Group UK London},
  doi       = {10.1038/s41567-018-0355-x},
  url       = {https://www.nature.com/articles/s41567-018-0355-x}
}

@article{Li.Science.2017,
  title     = "{Even-denominator fractional quantum Hall states in bilayer graphene}",
  author    = {Li, JIA and Tan, C and Chen, S and Zeng, Y and Taniguchi, T and Watanabe, K and Hone, J and Dean, CR},
  journal   = {Science},
  volume    = {358},
  number    = {6363},
  pages     = {648--652},
  year      = {2017},
  publisher = {American Association for the Advancement of Science},
  doi       = {10.1126/science.aao25},
  url       = {https://www.science.org/doi/full/10.1126/science.aao2521}
}

@article{Ki.NanoLett.2014,
  title     = "{Observation of even denominator fractional quantum Hall effect in suspended bilayer graphene}",
  author    = {Ki, Dong-Keun and Fal’ko, Vladimir I and Abanin, Dmitry A and Morpurgo, Alberto F},
  journal   = {Nano Lett.},
  volume    = {14},
  number    = {4},
  pages     = {2135--2139},
  year      = {2014},
  publisher = {ACS Publications},
  doi       = {10.1021/nl5003922},
  url       = {https://pubs.acs.org/doi/full/10.1021/nl5003922}
}

@article{Zibrov.Nature.2017,
  title     = "{Tunable interacting composite fermion phases in a half-filled bilayer-graphene Landau level}",
  author    = {Zibrov, Alexander A and Kometter, C and Zhou, H and Spanton, EM and Taniguchi, T and Watanabe, K and Zaletel, MP and Young, AF},
  journal   = {Nature},
  volume    = {549},
  number    = {7672},
  pages     = {360--364},
  year      = {2017},
  publisher = {Nature Publishing Group UK London},
  doi       = {10.1038/nature23893},
  url       = {https://www.nature.com/articles/nature23893#citeas}
}

@article{Liu.PRL.2014,
  title     = "{Fractional Quantum Hall Effect at $\ensuremath{\nu}=1/2$ in Hole Systems Confined to GaAs Quantum Wells}",
  author    = {Liu, Yang and Graninger, A. L. and Hasdemir, S. and Shayegan, M. and Pfeiffer, L. N. and West, K. W. and Baldwin, K. W. and Winkler, R.},
  journal   = {Phys. Rev. Lett.},
  volume    = {112},
  issue     = {4},
  pages     = {046804},
  numpages  = {5},
  year      = {2014},
  month     = {Jan},
  publisher = {American Physical Society},
  doi       = {10.1103/PhysRevLett.112.046804},
  url       = {https://link.aps.org/doi/10.1103/PhysRevLett.112.046804}
}

@article{Liu.PRB.2014,
  title     = "{Even-denominator fractional quantum Hall effect at a Landau level crossing}",
  author    = {Liu, Yang and Hasdemir, S. and Kamburov, D. and Graninger, A. L. and Shayegan, M. and Pfeiffer, L. N. and West, K. W. and Baldwin, K. W. and Winkler, R.},
  journal   = {Phys. Rev. B},
  volume    = {89},
  issue     = {16},
  pages     = {165313},
  numpages  = {6},
  year      = {2014},
  month     = {Apr},
  publisher = {American Physical Society},
  doi       = {10.1103/PhysRevB.89.165313},
  url       = {https://link.aps.org/doi/10.1103/PhysRevB.89.165313}
}

@article{Lay.PRB.1997,
  title     = "{One-component to two-component transition of the $\ensuremath{\nu}=2/3$ fractional quantum Hall effect in a wide quantum well induced by an in-plane magnetic field}",
  author    = {Lay, T. S. and Jungwirth, T. and Smr\ifmmode \check{c}\else \v{c}\fi{}ka, L. and Shayegan, M.},
  journal   = {Phys. Rev. B},
  volume    = {56},
  issue     = {12},
  pages     = {R7092--R7095},
  numpages  = {0},
  year      = {1997},
  month     = {Sep},
  publisher = {American Physical Society},
  doi       = {10.1103/PhysRevB.56.R7092},
  url       = {https://link.aps.org/doi/10.1103/PhysRevB.56.R7092}
}

@article{Shafayat.PRL.2023,
  title     = "{Valley-Tunable Even-Denominator Fractional Quantum Hall State in the Lowest Landau Level of an Anisotropic System}",
  author    = {Hossain, Md. Shafayat and Ma, Meng K. and Chung, Y. J. and Singh, S. K. and Gupta, A. and West, K. W. and Baldwin, K. W. and Pfeiffer, L. N. and Winkler, R. and Shayegan, M.},
  journal   = {Phys. Rev. Lett.},
  volume    = {130},
  issue     = {12},
  pages     = {126301},
  numpages  = {7},
  year      = {2023},
  month     = {Mar},
  publisher = {American Physical Society},
  doi       = {10.1103/PhysRevLett.130.126301},
  url       = {https://link.aps.org/doi/10.1103/PhysRevLett.130.126301}
}

@article{Huang.PRX.2022,
  title     = "{Valley Isospin Controlled Fractional Quantum Hall States in Bilayer Graphene}",
  author    = {Huang, Ke and Fu, Hailong and Hickey, Danielle Reifsnyder and Alem, Nasim and Lin, Xi and Watanabe, Kenji and Taniguchi, Takashi and Zhu, Jun},
  journal   = {Phys. Rev. X},
  volume    = {12},
  issue     = {3},
  pages     = {031019},
  numpages  = {14},
  year      = {2022},
  month     = {Jul},
  publisher = {American Physical Society},
  doi       = {10.1103/PhysRevX.12.031019},
  url       = {https://link.aps.org/doi/10.1103/PhysRevX.12.031019}
}

@article{Manoharan.PRL.1997,
  title     = "{Spontaneous Interlayer Charge Transfer near the Magnetic Quantum Limit}",
  author    = {Manoharan, H. C. and Suen, Y. W. and Lay, T. S. and Santos, M. B. and Shayegan, M.},
  journal   = {Phys. Rev. Lett.},
  volume    = {79},
  issue     = {14},
  pages     = {2722--2725},
  numpages  = {0},
  year      = {1997},
  month     = {Oct},
  publisher = {American Physical Society},
  doi       = {10.1103/PhysRevLett.79.2722},
  url       = {https://link.aps.org/doi/10.1103/PhysRevLett.79.2722}
}

@article{Pan.PRL.1999,
  title     = "{Exact Quantization of the Even-Denominator Fractional Quantum Hall State at $\mathit{\ensuremath{\nu}}\phantom{\rule{0ex}{0ex}}=\phantom{\rule{0ex}{0ex}}5/2$ Landau Level Filling Factor}",
  author    = {Pan, W. and Xia, J.-S. and Shvarts, V. and Adams, D. E. and Stormer, H. L. and Tsui, D. C. and Pfeiffer, L. N. and Baldwin, K. W. and West, K. W.},
  journal   = {Phys. Rev. Lett.},
  volume    = {83},
  issue     = {17},
  pages     = {3530--3533},
  numpages  = {0},
  year      = {1999},
  month     = {Oct},
  publisher = {American Physical Society},
  doi       = {10.1103/PhysRevLett.83.3530},
  url       = {https://link.aps.org/doi/10.1103/PhysRevLett.83.3530}
}

@article{Falson.Nat.Phys.2015,
  title     = "{Even-denominator fractional quantum Hall physics in ZnO}",
  author    = {Falson, J and Maryenko, D and Friess, B and Zhang, D and Kozuka, Y and Tsukazaki, A and Smet, JH and Kawasaki, M},
  journal   = {Nat. Phys.},
  volume    = {11},
  number    = {4},
  pages     = {347--351},
  year      = {2015},
  publisher = {Nature Publishing Group UK London},
  doi       = {https://doi.org/10.1038/nphys3259}
}

@article{Luhman.PRL.2008,
  title     = "{Observation of a Fractional Quantum Hall State at $\ensuremath{\nu}=1/4$ in a Wide GaAs Quantum Well}",
  author    = {Luhman, D. R. and Pan, W. and Tsui, D. C. and Pfeiffer, L. N. and Baldwin, K. W. and West, K. W.},
  journal   = {Phys. Rev. Lett.},
  volume    = {101},
  issue     = {26},
  pages     = {266804},
  numpages  = {4},
  year      = {2008},
  month     = {Dec},
  publisher = {American Physical Society},
  doi       = {10.1103/PhysRevLett.101.266804},
  url       = {https://link.aps.org/doi/10.1103/PhysRevLett.101.266804}
}

@article{Shabani2.PRL.2009,
  title     = "{Evidence for Developing Fractional Quantum Hall States at Even Denominator $1/2$ and $1/4$ Fillings in Asymmetric Wide Quantum Wells}",
  author    = {Shabani, J. and Gokmen, T. and Chiu, Y. T. and Shayegan, M.},
  journal   = {Phys. Rev. Lett.},
  volume    = {103},
  issue     = {25},
  pages     = {256802},
  numpages  = {4},
  year      = {2009},
  month     = {Dec},
  publisher = {American Physical Society},
  doi       = {10.1103/PhysRevLett.103.256802},
  url       = {https://link.aps.org/doi/10.1103/PhysRevLett.103.256802}
}

@article{Shabani.PRL.2009,
  title     = "{Correlated States of Electrons in Wide Quantum Wells at Low Fillings: The Role of Charge Distribution Symmetry}",
  author    = {Shabani, J. and Gokmen, T. and Shayegan, M.},
  journal   = {Phys. Rev. Lett.},
  volume    = {103},
  issue     = {4},
  pages     = {046805},
  numpages  = {4},
  year      = {2009},
  month     = {Jul},
  publisher = {American Physical Society},
  doi       = {10.1103/PhysRevLett.103.046805},
  url       = {https://link.aps.org/doi/10.1103/PhysRevLett.103.046805}
}

@article{Suen.PRB.1991,
  title     = "{Missing integral quantum Hall effect in a wide single quantum well}",
  author    = {Suen, Y. W. and Jo, J. and Santos, M. B. and Engel, L. W. and Hwang, S. W. and Shayegan, M.},
  journal   = {Phys. Rev. B},
  volume    = {44},
  issue     = {11},
  pages     = {5947--5950},
  numpages  = {0},
  year      = {1991},
  month     = {Sep},
  publisher = {American Physical Society},
  doi       = {10.1103/PhysRevB.44.5947},
  url       = {https://link.aps.org/doi/10.1103/PhysRevB.44.5947}
}

@inproceedings{Shayegan.Review.LesHouches.1999,
  title         = "{Electrons in a Flatland}",
  author        = {Shayegan, M},
  booktitle     = "{{1998 Les Houches Summer School, Session LXIX, Topological Aspects of Low Dimensional Systems}}",
  series        = {NATO Advanced Study Institute},
  doi           = {https://doi.org/10.1007/3-540-46637-1_1},
  editor        = {Comtet, A. and Jolic{\oe}ur, T. and Ouvry, S. and David, F},
  organization  = {Springer-Verlag},
  address       = {Berlin},
  year          = {1999},
  pages         = {1--51}
}

@book{Jain.composite.fermions.2007,
  title     = "{Composite Fermions}",
  author    = {Jain, Jainendra K},
  year      = {2007},
  doi       = {https://doi.org/10.1017/CBO9780511607561},
  publisher = {Cambridge University Press},
  address   = {Cambridge, England}
}

@article{Scarola.PRB.2001,
  title     = "{Phase diagram of bilayer composite fermion states}",
  author    = {Scarola, V. W. and Jain, J. K.},
  journal   = {Phys. Rev. B},
  volume    = {64},
  issue     = {8},
  pages     = {085313},
  numpages  = {10},
  year      = {2001},
  month     = {Aug},
  publisher = {American Physical Society},
  doi       = {10.1103/PhysRevB.64.085313},
  url       = {https://link.aps.org/doi/10.1103/PhysRevB.64.085313}
}

@article{Li.Nat.Phys.2019,
  title     = "{Pairing states of composite fermions in double-layer graphene}",
  author    = {Li, J. I. A. and Shi, Q and Zeng, Y and Watanabe, K and Taniguchi, T and Hone, J and Dean, CR},
  journal   = {Nat. Phys.},
  volume    = {15},
  number    = {9},
  pages     = {898--903},
  year      = {2019},
  publisher = {Nature Publishing Group UK London},
  doi       = {https://doi.org/10.1038/s41567-019-0547-z}
}

@article{Kumar.Nat.Comm.2025,
      title     = "{Quarter- and half-filled quantum Hall states and their competing interactions in bilayer graphene}", 
      author    = {Ravi Kumar and André Haug and Jehyun Kim and Misha Yutushui and Konstantin Khudiakov and Vishal Bhardwaj and Alexey Ilin and Kenji Watanabe and Takashi Taniguchi and David F. Mross and Yuval Ronen},
      journal   = {Nat. Comm.},
      volume    = {16},
      number    = {1},
      pages     = {7255},
      year      = {2025},
      publisher = {Nature Publishing Group UK London},
      doi       = {https://doi.org/10.1038/s41467-025-62650-9}
}

@article{Singh.PRL.2025,
  title     = "{Fractional Quantum Hall State at $\ensuremath{\nu}\text{}=\text{}1/2$ with Energy Gap Up to 6 K and Possible Transition from the One- to Two-Component State}",
  author    = {Singh, Siddharth Kumar and Wang, Chengyu and Gupta, Adbhut and Baldwin, Kirk W. and Pfeiffer, Loren N. and Shayegan, Mansour},
  journal   = {Phys. Rev. Lett.},
  volume    = {135},
  issue     = {24},
  pages     = {246603},
  numpages  = {8},
  year      = {2025},
  month     = {Dec},
  publisher = {American Physical Society},
  doi       = {10.1103/ywpx-qm7d},
  url       = {https://link.aps.org/doi/10.1103/ywpx-qm7d}
}

@article{Faugno.PRB.2020,
  title     = "{Theoretical phase diagram of two-component composite fermions in double-layer graphene}",
  author    = {Faugno, W. N. and Balram, Ajit C. and W\'ojs, A. and Jain, J. K.},
  journal   = {Phys. Rev. B},
  volume    = {101},
  issue     = {8},
  pages     = {085412},
  numpages  = {9},
  year      = {2020},
  month     = {Feb},
  publisher = {American Physical Society},
  doi       = {10.1103/PhysRevB.101.085412},
  url       = {https://link.aps.org/doi/10.1103/PhysRevB.101.085412}
}

@article{Pan.PRL.2003,
  title     = "{Fractional Quantum Hall Effect of Composite Fermions}",
  author    = {Pan, W. and Stormer, H. L. and Tsui, D. C. and Pfeiffer, L. N. and Baldwin, K. W. and West, K. W.},
  journal   = {Phys. Rev. Lett.},
  volume    = {90},
  issue     = {1},
  pages     = {016801},
  numpages  = {4},
  year      = {2003},
  month     = {Jan},
  publisher = {American Physical Society},
  doi       = {10.1103/PhysRevLett.90.016801},
  url       = {https://link.aps.org/doi/10.1103/PhysRevLett.90.016801}
}

@article{Faugno.PRL.2019,
  title     = "{Prediction of a Non-Abelian Fractional Quantum Hall State with $f$-Wave Pairing of Composite Fermions in Wide Quantum Wells}",
  author    = {Faugno, W. N. and Balram, Ajit C. and Barkeshli, Maissam and Jain, J. K.},
  journal   = {Phys. Rev. Lett.},
  volume    = {123},
  issue     = {1},
  pages     = {016802},
  numpages  = {6},
  year      = {2019},
  month     = {Jul},
  publisher = {American Physical Society},
  doi       = {10.1103/PhysRevLett.123.016802},
  url       = {https://link.aps.org/doi/10.1103/PhysRevLett.123.016802}
}

@article{Zibrov.Nat.Phys.2018,
  title     = "{Even-denominator fractional quantum Hall states at an isospin transition in monolayer graphene}",
  author    = {Zibrov, AA and Spanton, EM and Zhou, H and Kometter, C and Taniguchi, T and Watanabe, K and Young, AF},
  journal   = {Nat. Phys.},
  volume    = {14},
  number    = {9},
  pages     = {930--935},
  year      = {2018},
  publisher = {Nature Publishing Group UK London},
  doi       = {https://doi.org/10.1038/s41567-018-0190-0},
  url       = {https://www.nature.com/articles/s41567-018-0190-0#citeas}
}

@article{Chung.PRB.2022,
  title     = "{Understanding limits to mobility in ultrahigh-mobility GaAs two-dimensional electron systems: 100 million ${\mathrm{cm}}^{2}/\mathrm{Vs}$ and beyond}",
  author    = {Chung, Yoon Jang and Gupta, A. and Baldwin, K. W. and West, K. W. and Shayegan, M. and Pfeiffer, L. N.},
  journal   = {Phys. Rev. B},
  volume    = {106},
  issue     = {7},
  pages     = {075134},
  numpages  = {11},
  year      = {2022},
  month     = {Aug},
  publisher = {American Physical Society},
  doi       = {10.1103/PhysRevB.106.075134},
  url       = {https://link.aps.org/doi/10.1103/PhysRevB.106.075134}
}

@article{Zhao.PRL.2023,
  title     = "{Composite Fermion Pairing Induced by Landau Level Mixing}",
  author    = {Zhao, Tongzhou and Balram, Ajit C. and Jain, J. K.},
  journal   = {Phys. Rev. Lett.},
  volume    = {130},
  issue     = {18},
  pages     = {186302},
  numpages  = {7},
  year      = {2023},
  month     = {May},
  publisher = {American Physical Society},
  doi       = {10.1103/PhysRevLett.130.186302},
  url       = {https://link.aps.org/doi/10.1103/PhysRevLett.130.186302}
}

@inbook{Girvin.MacDonald.chapter,
author      = {Girvin, S. M. and MacDonald, A. H.},
publisher   = {John Wiley \& Sons, Ltd},
isbn        = {9783527617258},
title       = "{Multicomponent Quantum Hall Systems: The Sum of Their Parts and More}",
booktitle   = {Perspectives in Quantum Hall Effects},
editor      = {Sarma, S. D. and Pinczuk, A.},
chapter     = {5},
pages       = {161-224},
doi         = {https://doi.org/10.1002/9783527617258.ch5},
url         = {https://onlinelibrary.wiley.com/doi/abs/10.1002/9783527617258.ch5},
year        = {1996}
}

@article{Peterson.PRB.2015,
  title     = "{Abelian and non-Abelian states in $\ensuremath{\nu}=2/3$ bilayer fractional quantum Hall systems}",
  author    = {Peterson, Michael R. and Wu, Yang-Le and Cheng, Meng and Barkeshli, Maissam and Wang, Zhenghan and Das Sarma, Sankar},
  journal   = {Phys. Rev. B},
  volume    = {92},
  issue     = {3},
  pages     = {035103},
  numpages  = {7},
  year      = {2015},
  month     = {Jul},
  publisher = {American Physical Society},
  doi       = {10.1103/PhysRevB.92.035103},
  url       = {https://link.aps.org/doi/10.1103/PhysRevB.92.035103}
}

@article{Halperin.Lee.Read.PRB.1993,
  title     = "{Theory of the half-filled Landau level}",
  author    = {Halperin, B. I. and Lee, Patrick A. and Read, Nicholas},
  journal   = {Phys. Rev. B},
  volume    = {47},
  issue     = {12},
  pages     = {7312--7343},
  numpages  = {0},
  year      = {1993},
  month     = {Mar},
  publisher = {American Physical Society},
  doi       = {10.1103/PhysRevB.47.7312},
  url       = {https://link.aps.org/doi/10.1103/PhysRevB.47.7312}
}

@article{Willett.PRL.1993,
  title     = "{Experimental demonstration of a Fermi surface at one-half filling of the lowest Landau level}",
  author    = {Willett, R. L. and Ruel, R. R. and West, K. W. and Pfeiffer, L. N.},
  journal   = {Phys. Rev. Lett.},
  volume    = {71},
  issue     = {23},
  pages     = {3846--3849},
  numpages  = {0},
  year      = {1993},
  month     = {Dec},
  publisher = {American Physical Society},
  doi       = {10.1103/PhysRevLett.71.3846},
  url       = {https://link.aps.org/doi/10.1103/PhysRevLett.71.3846}
}

@article{Park.PRB.1998,
  title     = "{Possibility of p-wave pairing of composite fermions at $\ensuremath{\nu}=\frac{1}{2}$}",
  author    = {Park, K. and Melik-Alaverdian, V. and Bonesteel, N. E. and Jain, J. K.},
  journal   = {Phys. Rev. B},
  volume    = {58},
  issue     = {16},
  pages     = {R10167--R10170},
  numpages  = {0},
  year      = {1998},
  month     = {Oct},
  publisher = {American Physical Society},
  doi       = {10.1103/PhysRevB.58.R10167},
  url       = {https://link.aps.org/doi/10.1103/PhysRevB.58.R10167}
}

@article{Goldman.PRL.1988,
  title     = "{Evidence for the Fractional Quantum Hall State at $\ensuremath{\nu}=\frac{1}{7}$}",
  author    = {Goldman, V. J. and Shayegan, M. and Tsui, D. C.},
  journal   = {Phys. Rev. Lett.},
  volume    = {61},
  issue     = {7},
  pages     = {881--884},
  numpages  = {0},
  year      = {1988},
  month     = {Aug},
  publisher = {American Physical Society},
  doi       = {10.1103/PhysRevLett.61.881},
  url       = {https://link.aps.org/doi/10.1103/PhysRevLett.61.881}
}

@article{Samkharadze.PRB.2015,
  title     = "{Observation of incompressibility at $\ensuremath{\nu}=4/11$ and $\ensuremath{\nu}=5/13$}",
  author    = {Samkharadze, N. and Arnold, I. and Pfeiffer, L. N. and West, K. W. and Cs\'athy, G. A.},
  journal   = {Phys. Rev. B},
  volume    = {91},
  issue     = {8},
  pages     = {081109},
  numpages  = {4},
  year      = {2015},
  month     = {Feb},
  publisher = {American Physical Society},
  doi       = {10.1103/PhysRevB.91.081109},
  url       = {https://link.aps.org/doi/10.1103/PhysRevB.91.081109}
}

@article{Sajoto.PRB.1990,
  title     = "{Fractional quantum Hall effect in very-low-density GaAs/${\mathrm{Al}}_{\mathit{x}}$${\mathrm{Ga}}_{1\mathrm{\ensuremath{-}}\mathit{x}}$As heterostructures}",
  author    = {Sajoto, T. and Suen, Y. W. and Engel, L. W. and Santos, M. B. and Shayegan, M.},
  journal   = {Phys. Rev. B},
  volume    = {41},
  issue     = {12},
  pages     = {8449--8460},
  numpages  = {0},
  year      = {1990},
  month     = {Apr},
  publisher = {American Physical Society},
  doi       = {10.1103/PhysRevB.41.8449},
  url       = {https://link.aps.org/doi/10.1103/PhysRevB.41.8449}
}

@article{Huang.NatComm.2024,
  title     = "{Evidence for Topological Protection Derived from Six-Flux Composite Fermions}",
  author    = {Huang, Haoyun and Hussain, Waseem and Myers, SA and Pfeiffer, LN and West, KW and Baldwin, KW and Cs{\'a}thy, GA},
  journal   = {Nat. Comm.},
  volume    = {15},
  number    = {1},
  pages     = {1461},
  year      = {2024},
  publisher = {Nature Publishing Group UK London},
  doi       = {10.1038/s41467-024-45860-5},
  url       = {https://www.nature.com/articles/s41467-024-45860-5}
}

@misc{SM.2025,
    note = "{See Supplemental Material for a detailed discussion on the 1C to 2C transition of the 2DES for $2>\nu>1$. }"
}

@misc{twothird.fourseventh,
    note = "{Note however that the 1C to 2C transitions do not all occure at the same $\alpha \simeq 0.064$. For example, the transition happens at $\alpha \simeq 0.073$ and 0.060 for $\nu = 2/3$ and 4/7 FQHSs, respectively.}"
}

@unpublished{Bell.Preprint.2025,
      title         = "{High-order two-component fractional quantum Hall states around filling factor $\nu = 1$}", 
      author        = {E. Bell and K. W. Baldwin and L. N. Pfeiffer and K. W. West and M. A. Zudov},
      year          = {2025},
      eprint        = {2512.04050},
      archivePrefix = {arXiv},
      primaryClass  = {cond-mat.mes-hall},
      url           = {https://arxiv.org/abs/2512.04050}, 
}

@article{Zhang.Nature.2025,
  title     = "{Excitons in the fractional quantum Hall effect}",
  author    = {Zhang, Naiyuan J and Nguyen, Ron Q and Batra, Navketan and Liu, Xiaoxue and Watanabe, Kenji and Taniguchi, Takashi and Feldman, DE and Li, JIA},
  journal   = {Nature},
  volume    = {637},
  number    = {8045},
  pages     = {327--332},
  year      = {2025},
  publisher = {Nature Publishing Group UK London}, 
  doi       = {https://doi.org/10.1038/s41586-024-08274-3}
}

@misc{note.2C.4_5.FQHS,
    note = "{Alternate possibilities for the 4/5 and 4/7 FQHSs are the (2/3, 2/3 $|$1) and (2/5, 2/5 $|$1) states, respectively. Similarly, the 11/15 FQHS can also be described by the (3/5, 5/9 $|$1) state. The stability of the different competing states is strongly dependent on 2DES parameters such as $\alpha$, d/$\ell_B$ and $\lambda/\ell_B$. Further experimental investigations sensitive to the topological order, and numerical calculations such as those of Ref. \protect{\cite{Faugno.PRB.2020}} can shed more light.}"
}

@article{Eisenstein.Annu.Review.2014,
  title     = "{Exciton condensation in bilayer quantum Hall systems}",
  author    = {Eisenstein, JP},
  journal   = {Annu. Rev. Condens. Matter Phys.},
  volume    = {5},
  number    = {1},
  pages     = {159--181},
  year      = {2014},
  publisher = {Annual Reviews},
  doi       = {https://doi.org/10.1146/annurev-conmatphys-031113-133832}
}

@article{Murphy.PRL.1994,
  title     = "{Many-body integer quantum Hall effect: Evidence for new phase transitions}",
  author    = {Murphy, S. Q. and Eisenstein, J. P. and Boebinger, G. S. and Pfeiffer, L. N. and West, K. W.},
  journal   = {Phys. Rev. Lett.},
  volume    = {72},
  issue     = {5},
  pages     = {728--731},
  numpages  = {0},
  year      = {1994},
  month     = {Jan},
  publisher = {American Physical Society},
  doi       = {10.1103/PhysRevLett.72.728},
  url       = {https://link.aps.org/doi/10.1103/PhysRevLett.72.728}
}

@article{Yang.PRB.1996,
  title     = "{Spontaneous interlayer coherence in double-layer quantum Hall systems: Symmetry-breaking interactions, in-plane fields, and phase solitons}",
  author    = {Yang, Kun and Moon, K. and Belkhir, Lotfi and Mori, H. and Girvin, S. M. and MacDonald, A. H. and Zheng, L. and Yoshioka, D.},
  journal   = {Phys. Rev. B},
  volume    = {54},
  issue     = {16},
  pages     = {11644--11658},
  numpages  = {0},
  year      = {1996},
  month     = {Oct},
  publisher = {American Physical Society},
  doi       = {10.1103/PhysRevB.54.11644},
  url       = {https://link.aps.org/doi/10.1103/PhysRevB.54.11644}
}

@article{Hu.Nat.Phys.2025,
  title     = "{High-resolution tunnelling spectroscopy of fractional quantum Hall states}",
  author    = {Hu, Yuwen and Tsui, Yen-Chen and He, Minhao and Kamber, Umut and Wang, Taige and Mohammadi, Amir S and Watanabe, Kenji and Taniguchi, Takashi and Papi{\'c}, Zlatko and Zaletel, Michael P and others},
  journal   = {Nat. Phys.},
  volume    = {21}, 
  number    = {5},
  pages     = {716--723},
  year      = {2025},
  month     = {March},
  publisher = {Nature Publishing Group UK London},
  doi       = {10.1038/s41567-025-02830-y},
  url       = {https://www.nature.com/articles/s41567-025-02830-y}
}

@article{Jain.PRB.1989,
  title     = "{Incompressible quantum Hall states}",
  author    = {Jain, J. K.},
  journal   = {Phys. Rev. B},
  volume    = {40},
  issue     = {11},
  pages     = {8079--8082},
  numpages  = {0},
  year      = {1989},
  month     = {Oct},
  publisher = {American Physical Society},
  doi       = {10.1103/PhysRevB.40.8079},
  url       = {https://link.aps.org/doi/10.1103/PhysRevB.40.8079}
}

@article{Laughlin.PRL.1983,
  title     = "{Anomalous Quantum Hall Effect: An Incompressible Quantum Fluid with Fractionally Charged Excitations}",
  author    = {Laughlin, R. B.},
  journal   = {Phys. Rev. Lett.},
  volume    = {50},
  issue     = {18},
  pages     = {1395--1398},
  numpages  = {0},
  year      = {1983},
  month     = {May},
  publisher = {American Physical Society},
  doi       = {10.1103/PhysRevLett.50.1395},
  url       = {https://link.aps.org/doi/10.1103/PhysRevLett.50.1395}
}

@article{Jain.PRL.1989,
  title     = "{Composite-fermion approach for the fractional quantum Hall effect}",
  author    = {Jain, J. K.},
  journal   = {Phys. Rev. Lett.},
  volume    = {63},
  issue     = {2},
  pages     = {199--202},
  numpages  = {0},
  year      = {1989},
  month     = {Jul},
  publisher = {American Physical Society},
  doi       = {10.1103/PhysRevLett.63.199},
  url       = {https://link.aps.org/doi/10.1103/PhysRevLett.63.199}
}

@article{Xia.PRL.2004,
  title     = "{Electron Correlation in the Second Landau Level: A Competition Between Many Nearly Degenerate Quantum Phases}",
  author    = {Xia, J. S. and Pan, W. and Vicente, C. L. and Adams, E. D. and Sullivan, N. S. and Stormer, H. L. and Tsui, D. C. and Pfeiffer, L. N. and Baldwin, K. W. and West, K. W.},
  journal   = {Phys. Rev. Lett.},
  volume    = {93},
  issue     = {17},
  pages     = {176809},
  numpages  = {4},
  year      = {2004},
  month     = {Oct},
  publisher = {American Physical Society},
  doi       = {10.1103/PhysRevLett.93.176809},
  url       = {https://link.aps.org/doi/10.1103/PhysRevLett.93.176809}
}

@article{Chen.Nat.Comm.2024,
  title     = "{Tunable even-and odd-denominator fractional quantum Hall states in trilayer graphene}",
  author    = {Chen, Yiwei and Huang, Yan and Li, Qingxin and Tong, Bingbing and Kuang, Guangli and Xi, Chuanying and Watanabe, Kenji and Taniguchi, Takashi and Liu, Guangtong and Zhu, Zheng and others},
  journal   = {Nat. Comm.},
  volume    = {15},
  number    = {1},
  pages     = {6236},
  year      = {2024},
  publisher = {Nature Publishing Group UK London}, 
  doi       = {10.1038/s41467-024-50589-2},
  url       = {https://www.nature.com/articles/s41467-024-50589-2}
}

@unpublished{Chanda.Preprint.2025,
      title         = "{Even denominator fractional quantum Hall states in the zeroth Landau level of monolayer-like band of ABA trilayer graphene}", 
      author        = {Tanima Chanda and Simrandeep Kaur and Harsimran Singh and Kenji Watanabe and Takashi Taniguchi and Manish Jain and Udit Khanna and Ajit C. Balram and Aveek Bid},
      year          = {2025},
      eprint        = {2502.06245},
      archivePrefix = {arXiv},
      primaryClass  = {cond-mat.mes-hall},
      url           = {https://arxiv.org/abs/2502.06245}, 
      doi           = {10.48550/arXiv.2502.06245}
}

@misc{spin.polarization.CFs,
    note = "{We emphasize that in the range of densities of interest in our 72.5-nm-wide sample, the composite fermions around \protect{$\nu = 3/2$} and 1/2 are fully spin polarized. Therefore, the only other relevant energy scale for the 1C to 2C transition is \protect{$\Delta_{SAS}$}. Refer to \cite{Liu2.PRB.2014} for a phase diagram of the spin polarization of composite fermions.}"
}

\end{document}


\title{Supplemental Material for ``Observation of even-denominator fractional quantum Hall states at $\nu = 3/4$ and 5/4 in the lowest Landau level"}
\date{\today}
    
\author{Siddharth Kumar Singh} 
\affiliation{Department of Physics, Columbia University, New York, New York 10027, USA}
\author{Chengyu Wang} 
\affiliation{Department of Electrical and Computer Engineering, Princeton University, Princeton, New Jersey 08544, USA}
\author{Adbhut Gupta}
\affiliation{Department of Electrical and Computer Engineering, Princeton University, Princeton, New Jersey 08544, USA}
\author{Kirk W. Baldwin}
\affiliation{Department of Electrical and Computer Engineering, Princeton University, Princeton, New Jersey 08544, USA}
\author{Loren N. Pfeiffer}
\affiliation{Department of Electrical and Computer Engineering, Princeton University, Princeton, New Jersey 08544, USA}
\author{Mansour Shayegan}
\affiliation{Department of Electrical and Computer Engineering, Princeton University, Princeton, New Jersey 08544, USA}

\let\oldmaketitle\maketitle
\renewcommand\maketitle{{\bfseries\boldmath\oldmaketitle}}

\setcounter{figure}{0}
\renewcommand{\figurename}{FIG.}
\renewcommand{\thefigure}{S\arabic{figure}}
\renewcommand{\thesection}{S\arabic{section}}
\maketitle

\clearpage
\tableofcontents
\clearpage

\section{One-component to two-component transition in the $N = 0$ Landau level}

We discuss here the one-component (1C) to two-component (2C) transition of the 2DES in our 72.5-nm-wide GaAs quantum well (QW). This 1C to 2C transition of the 2DES is inferred from the evolution of the correlated states at $\nu < 2$ such as fractional quantum Hall states (FQHSs) and insulating phase. The 1C to 2C transition of 2DESs confined to wide GaAs QWs for $\nu < 1$ is discussed in detail in Refs. \cite{Manoharan.PRL.1996, Singh.PRL.2025}.

We capture coarsely the evolution of the nature of electronic correlations in the 2DES through the data presented in Figs. \hyperlink{app1-1}{S1}-\hyperlink{app1-3}{S3}. Figure \hyperlink{app1-1}{S1(a)} shows the characteristic longitudinal ($R_{xx}$) and Hall ($R_{xy}$) resistances in the \textit{low} density regime, at $n = 1.16$ {(in units of $10^{11}\ \rm{cm^{-2}}$ which we use throughout this Supplemental Material)}. The 1C nature of the 2DES is highlighted by the odd-denominator Jain FQHSs which are observed in single-layer 2DESs \cite{Jain.composite.fermions.2007}. For example, FQHSs are observed at $\nu = 2/5$, 3/7, 4/9,... on the high-field side of $\nu = 1/2$, and at $\nu = 2/3$, 3/5, 4/7,... on the low-field side of $\nu = 1/2$. These FQHSs become weaker as one approaches $\nu = 1/2$, as indicated by the strengths of the $R_{xx}$ minima. This behavior is typical of Jain FQHSs belonging to the 2-flux composite fermion (CF) sequence given by $\nu = \nu_{CF}/(2\nu_{CF}\pm1)$, where $\nu_{CF} = 1$, 2, 3,.. denotes the CF filling. Figure \hyperlink{app1-1}{S1(b)} shows an expanded view of $R_{xx}$ and $R_{xy}$ for $2>\nu>1$. The odd-denominator FQHSs flanking $\nu = 3/2$ also fall in the 2-flux CF sequence of Jain FQHSs. The odd-denominator FQHSs around $\nu = 3/2$ and 1/2 occur in the $N = 0$ LL of the antisymmetric and symmetric subbands, respectively. 

Odd-denominator FQHSs are also observed at $\nu = 9/7$ and 6/5 flanking the compressible state at $\nu = 5/4$. Similarly, odd-denominator FQHSs are observed at $\nu = 4/5$, 7/9, 8/11, and 5/7 flanking the compressible state at $\nu = 3/4$ [Fig. \hyperlink{app1-1}{S1(c)}]. These FQHSs are the 4-flux CF Jain FQHSs, and indicate the existence of a CF Fermi sea at $\nu = 5/4$ and 3/4. The odd-denominator FQHSs discussed thus far are all observed in single-layer 2DESs \cite{Jain.composite.fermions.2007, Willett.PRL.1987, Pan.PRL.2003, Chung.NatMater.2021} indicating that, despite the bilayer charge distribution, the ground state of the 2DES is dominated by its intralayer Couloumb interaction.

Figure \hyperlink{app1-2}{S2(a)} shows the characteristic $R_{xx}$ and $R_{xy}$ trace at $n = 1.59$. We refer to this density as the \textit{intermediate} density, and in this regime the 2DES is 2C over a large range of $\nu<1$. This is indicated by the weakening (and eventual disappearance) of the \textit{odd-numerator} FQHSs at $\nu = 7/9$, 5/7 and 3/5, while the \textit{even-numerator} FQHSs at $\nu = 2/3$ and 4/7 are robust as 2C, bilayer FQHSs; see Ref. \cite{Singh.PRL.2025}. The emergence of an insulating phase on either side of $\nu = 1/2$ is attirbuted to a pinned, bilayer Wigner crystal \cite{Manoharan.PRL.1996, Hatke.PRB.2017}. Figure \hyperlink{app1-2}{S2(c)} shows an expanded view of $R_{xx}$ and $R_{xy}$ around $\nu = 3/4$. There is a strong FQHS at $\nu = 3/4$. Also, clear signatures of FQHSs at $\nu = 11/15$ and 29/35 are observed; these are spontaneously-imbalanced, 2C, bilayer FQHSs \cite{Manoharan.PRL.1997}.The FQHSs at $\nu = 3/4$, 11/15 and 29/35 do not have analogs in single-layer 2DESs, and signal the enhanced role of the bilayer charge distribution of the 2DES and smaller $\Delta_{SAS}$. In contrast, the 2DES continues to be 1C for $2 > \nu > 1$ as clearly demonstrated in Fig. \hyperlink{app1-2}{S2(b)} which holds a stark similarity to the data presented in Fig. \hyperlink{app1-1}{S1(b)}. 

Finally, we present data in the high-density regime at $n = 1.83$ in Fig. \hyperlink{app1-3}{S3(a)}. The 2DES now shows 2C features for $2>\nu$ also. \textit{Even-numerator} FQHSs at $\nu = 2/3$ and 4/7 continue to be strong whereas the \textit{odd-numerator} FQHSs at $\nu = 7/9$ and 5/7 have disappeared. Figures \hyperlink{app1-3}{S3(b,c)} show the expanded views for $2>\nu>1$ and $1>\nu$, respectively. Even-numerator FQHSs at $\nu = 4/3$ and 6/5 remain robust, while odd-numerator FQHSs are weakened (e.g., at $\nu = 7/5$ and 9/7). Even-denominator FQHSs now emerge at $\nu = 3/2$ and 5/4 [Fig. \hyperlink{app1-3}{S3(b)}], together with several odd-denominator FQHSs at $\nu = 29/21$ and 19/15 [Fig. \hyperlink{app1-3}{S3(b)}], and $\nu = 55/63$, and 29/35 [Fig. \hyperlink{app1-3}{S3(c)}]. The $\nu = 6/7$ FQHS is a 2C, bilayer FQHS with 3/7 filling of each layer. The other FQHSs are the spontaneously-imbalanced, bilayer FQHSs with unequal layer fillings. We emphasize that the even-denominator FQHSs at $\nu = 3/2$, 5/4 and 3/4, and the spontaneously-imbalanced, odd-denominator FQHSs are not observed in single-layer 2DESs. 

\section{Phase diagrams of the 2DES in our 72.5-nm-wide GaAs quantum well}

The $R_{xx}$ traces for $0.98\lesssim n \lesssim 1.83$ are compiled together to construct a color-scale phase diagram of the 2DES in our 72.5-nm-wide QW, shown in Fig. \hyperlink{pd_n}{S4}. The data are plotted as a function of $1/\nu$ as the x-axis, and $n$ as the y-axis; the tick marks denote the expected positions of FQHSs at different $\nu$ for simplicity. {The white curve is a guide-to-the-eye, and is determined only from the the 1C to 2C transitions of the prominent, odd-denominator FQHSs for $\nu < 1$.} For $\nu < 1/2$, a bilayer Wigner crystal emerges as a very-high-resistance insulating phase, and becomes the ground state, wiping out all the Jain FQHSs and the 8/17 daughter state of the 1/2 FQHS. For $\nu > 1/2$, we observe clear transitions at even-numerator FQHS $\nu = 4/7$ and the destabilization of the odd-numerator FQHSs at $\nu = 7/9$, 5/7, 3/5, {and 7/13 daughter state of the 1/2 FQHS}. Clearly, the 1C to 2C transitions at any given $\nu$ occur at slightly different \textit{n}. For $2>\nu>1$, green squares are used to mark the emergence, as the density is raised, of the $\nu = 3/2$, 29/21, 19/15 and 5/4 FQHSs, and the 1C to 2C transitions of the even-numerator FQHSs at $\nu = 4/3$ and 6/5. {Note that at the highest densities, the integer QHS at $\nu = 2$ becomes extremely strong, and its wide $R_{xx}$ minimum engulfs the $\nu = 5/3$ and 8/5 FQHSs; this can be seen clearly in Fig. \hyperlink{app1-3}{S3(a)}}.

A natural question arises as to why the 1C to 2C transitions occur at different densities for different $\nu$. We try to shed some light on this by plotting the data in Fig. \hyperlink{pd_alpha}{S5} as a function of the parameter $\alpha = \Delta_{SAS}/(e^2/4\pi\epsilon\epsilon_0\ell_B)$ as the y-axis, where $\Delta_{SAS}$ is the subband separation between the two lowest electric subbands of the wide QW, $\epsilon\simeq13$ is the dielectric constant of GaAs and $\ell_B = \sqrt{\hbar/eB}$ is the magnetic length. The parameter $\alpha$ is the interlayer tunneling normalized to the Coulomb energy. We note that when we plot the data in this fashion, the 1C-2C transitions for $\nu<1$ occur near the same value of $\alpha \simeq 0.064$ \cite{twothird.fourseventh}. This highlights the prominent role of the interlayer tunneling in governing the energetics of the 2DES at low fillings. However, for $\nu > 1$, we note that the 1C to 2C transitions are shifted to larger values of $\alpha$. This is discussed in the next section. 

\section{Evolution of fractional quantum Hall states in the $d/\ell_B$ vs $\alpha$ phase space}

\begin{table}[t!]
\hypertarget{tab1}{}
\centering
    \begin{tabular}{wc{0.1\columnwidth}|wc{0.2\columnwidth} wc{0.2\columnwidth} wc{0.2\columnwidth} wc{0.2\columnwidth}}
        \hline\hline
         &\boldsymbol{$\nu$} & \boldsymbol{$\alpha$} & \boldsymbol{$d/\ell_B$} & \boldsymbol{$\lambda/\ell_B$}\\
         \cline{2-5}
         \multirow{6}{0.1\columnwidth}{$\nu>1$}&5/3 & 0.087 & 4.0 & 1.5\\
         &\color{red}3/2 & \color{red}0.079 & \color{red}4.3 & \color{red}1.5\\
         &\color{blue}29/21 & \color{blue}0.072 & \color{blue}4.6 & \color{blue}1.6\\\cline{2-5}
         &\color{blue}19/15 & \color{blue}0.069 & \color{blue}4.8 & \color{blue}1.7\\
         &\color{red}5/4 & \color{red}0.068 & \color{red}4.8 & \color{red}1.7\\
         &6/5 & 0.071 & 4.8 & 1.7\\\cline{1-5}
         \multirow{4}{0.1\columnwidth}{$\nu<1$}&7/9 & 0.065 & 5.7 & 2.1\\
         &\color{red}3/4 & \color{red}0.069 & \color{red}5.6 & \color{red}2.1\\
         &\color{blue}11/15 & \color{blue}0.067 & \color{blue}5.7 & \color{blue}2.2\\
         &5/7 & 0.063 & 5.9 & 2.2\\
         \hline\hline
    \end{tabular}
    \caption{Values of the parameters $\alpha,\ d/\ell_B$ and $\lambda/\ell_B$ at which some of the 2C FQHSs emerge, or 1C FQHSs (such as those at $\nu = 5/3$, 7/9 and 5/7) disappear, or undergo a 1C to 2C transition (e.g., at $\nu = 6/5$).}
\end{table}

We highlight that while the 1C to 2C transitions occur around $\alpha \simeq 0.064$ for $\nu < 1$, the same does not hold true for $2 > \nu > 1$. This could be related to the enhanced role of small $d/\ell_B$ and $\lambda/\ell_B$ at high fillings, depicted in Fig. \hyperlink{doverlb}{S6} and Table \hyperlink{tab1}{S1}. The parameters $d$ and $\lambda$ are denoted in the center panel of Fig. 1(b) of the main text; $d$ is the interlayer distance and $\lambda$, defined as the full-width-at-half maximum of the individual layers, is a measure of the layer thicknesses. 

The discussion regarding the 1C to 2C transition of the 2DES is more complex than the phase diagrams of Figs. \hyperlink{pd_n}{S4} and \hyperlink{pd_alpha}{S5} can capture accurately. This is because the 2DES in wide GaAs QWs has several competing energy scales in the form of interlayer tunneling $(\alpha)$, interlayer $(\sim1/d)$ and intralayer $(\sim1/\ell_B)$ Coulomb interaction, and the softening of the Coulomb repulsion because of finite layer thickness $(\lambda)$ \cite{Shabani.PRB.2013, Singh.PRL.2025}. {Table \hyperlink{tab1}{I} lists the values of $\alpha$, $d/\ell_B$ and $\lambda/\ell_B$ at specific $\nu$ where (i) 2C FQHSs emerge, (ii) FQHSs undergo their 1C to 2C transition, and (iii) 1C FQHSs disappear.} Notice that the 3/2 FQHS, which is likely a 2C FQHS, emerges at larger value of $\alpha\simeq0.079$ compared to the 5/4 and 3/4 FQHSs ($\alpha \simeq 0.068$ and 0.069, respectively). This can potentially be explained by the smaller $d/\ell_B \simeq 4.3$ when the 3/2 FQHS emerges as opposed to the larger $d/\ell_B \simeq 4.8$ and 5.6 when the 5/4 and 3/4 FQHSs emerge. 

The trajectories of the 1/2, 3/4, 5/4 and 3/2 ground states in the $d/\ell_B$ vs $\alpha$ phase space are shown in Fig. \hyperlink{doverlb}{S6}. Also plotted are the trajectories of the nearby odd-denominator FQHSs where clear 1C to 2C transitions are observed. Solid symbols denote the presence of a FQHS (or developing FQHS) and open symbols denote the absence of a FQHS at the denoted $\nu$. The disappearance of the \textit{odd-denominator} FQHSs are used to infer a 1C to 2C phase boundary, as depicted by a dashed black curve, a guide-to-the-eye. We note that the 1/2 FQHS is stable over a wide range of $\alpha$ across the 1C to 2C phase boundary, and at large values of $d/\ell_B$. It is suggestive that the 1/2 FQHS undergoes a 1C to 2C transition \cite{Singh.PRL.2025}. However, the 3/4, 5/4 and 3/2 FQHSs emerge very close to the 1C to 2C phase boundary. In addition, it is remarakable that the 1C to 2C phase boundary bends towards larger values of $\alpha$ as determined by the 1C to 2C transitions of the odd-denominator FQHSs; this occurs as $d/\ell_B\lesssim5$. This trend stays true also for the even-denominator 3/2 and 5/4 FQHSs, highlighting the prominent role of $d/\ell_B$ at smaller values and provides a potential explanation for the emergence of the 3/2 FQHS at a rather large value of $\alpha$ \cite{Girvin.MacDonald.chapter, Scarola.PRB.2001, Faugno.PRB.2020}.

\section{Anomalous evolution at $\nu = 5/4$ in the $N = 0$ Landau level}

Figure \hyperlink{fivequarterevolution}{S7} shows $R_{xx}$ and $dR_{xy}/dB$ traces around $\nu = 3/2$ and 5/4 for selected $n$. The ground state at $\nu = 3/2$ evolves from a compressible composite fermion Fermi sea to a robust FQHS. At $\nu = 5/4$, the 2DES is compressible for $n=1.41$, 1.63, and 1.73. A robust 2C FQHS is observed at $\nu = 5/4$ for $n=1.83$. An anomalous, weak minimum is observed for $n = 1.65$, 1.67, and this $R_{xx}$ minimum turns into a shoulder for $n=1.69$. Also shown on top of Fig. \hyperlink{fivequarterevolution}{S7} are $dR_{xy}/dB$ traces for $n = 1.41$, 1.67, and 1.83. The behavior of $dR_{xy}/dB$ corroborates our observations based on the $R_{xx}$ data. $dR_{xy}/dB$ is constant at $\nu = 5/4$ for $n=1.41$ denoting a compressible state, and dips below the classical Hall value for $n=1.67$, suggesting a developing FQHS. For $n = 1.83$, $dR_{xy}/dB$ is constant and vanishes, as is expected for a robust FQHS. The $R_{xx}$ minimum, as well as the inflection in $dR_{xy}/dB$ at $\nu = 5/4$ for $n=1.67$, suggests a developing FQHS.

Figure \hyperlink{fivequarter1C}{S8} shows the temperature dependence of the weak $R_{xx}$ minimum at $\nu = 5/4$ for $n = 1.67$. The $R_{xx}$ minimum vanishes quickly with temperature suggesting its fragile origin. The behavior is also consistent with a many-body origin. Possibility of a weak 1C 5/4 FQHS is discussed in the main text. This state is akin to the 1/4 FQHS observed in wide GaAs QWs \cite{Shabani.PRL.2009, Shabani2.PRL.2009, Faugno.PRL.2019}, but now stabilized in the $N = 0$ Landau level of the antisymmetric subband. 

\bibliography{references_arxiv.bib}
\newpage
\begin{figure*}[b!]
\hypertarget{app1-1}{}
\includegraphics[width=1\columnwidth]{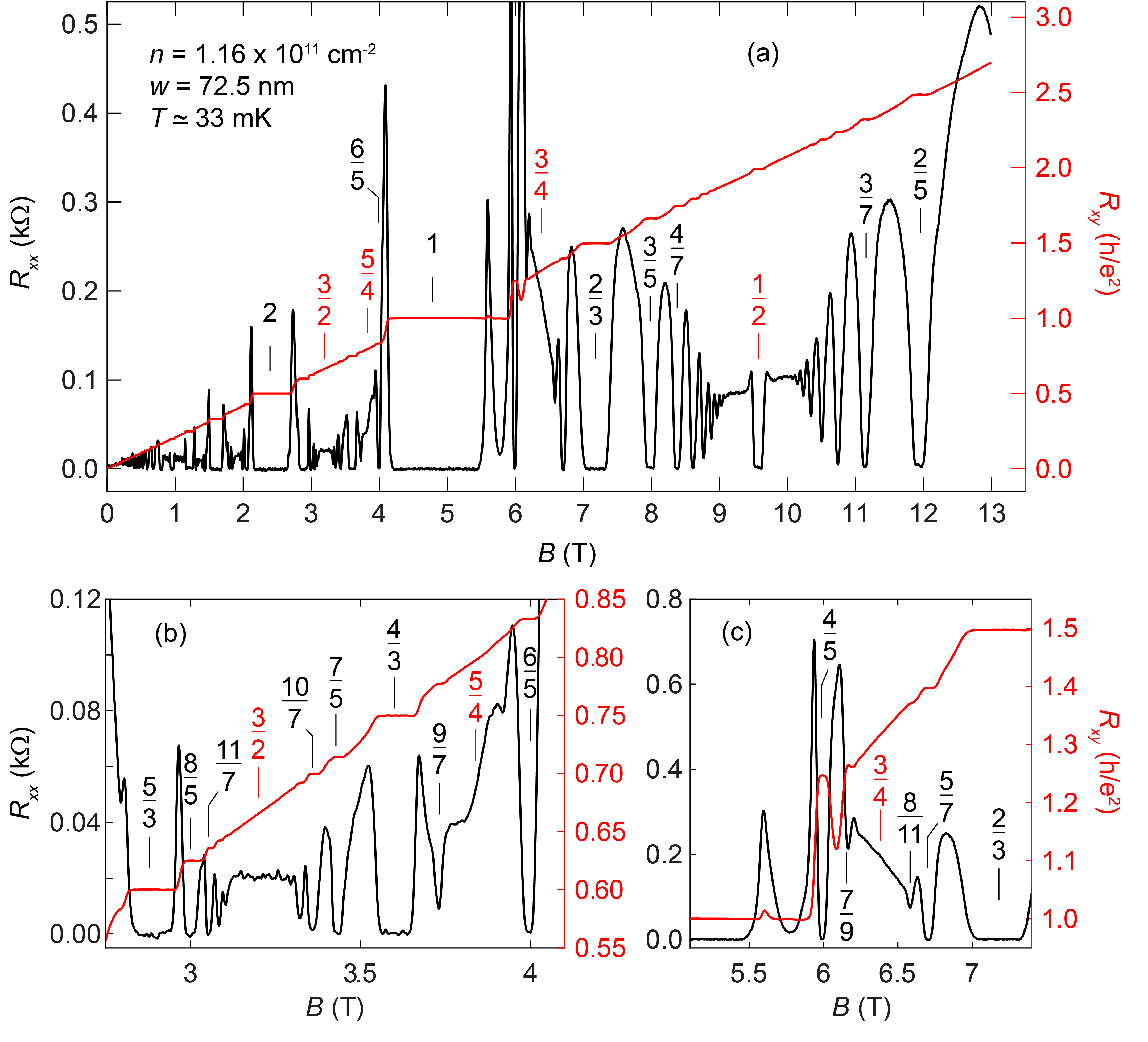}
\caption{(a) Full-field ${R_{xx}}$ and ${R_{xy}}$ traces at ${n = 1.16}\times10^{11}\ \rm{cm}^{-2}$. Expanded views of $R_{xx}$ and $R_{xy}$ around (b) $\nu = 3/2$ and 5/4, and (c) $\nu = 3/4$. $R_{xx}$ shows clear 1C behavior based on the observation of odd-denominator Jain FQHSs for $\nu \leq 2$. Vertical lines denote the expected positions of $\nu = 3/2$, 5/4, 3/4 and 1/2.}
\end{figure*}

\begin{figure*}[t!]
\hypertarget{app1-2}{}
\includegraphics[width=1\columnwidth]{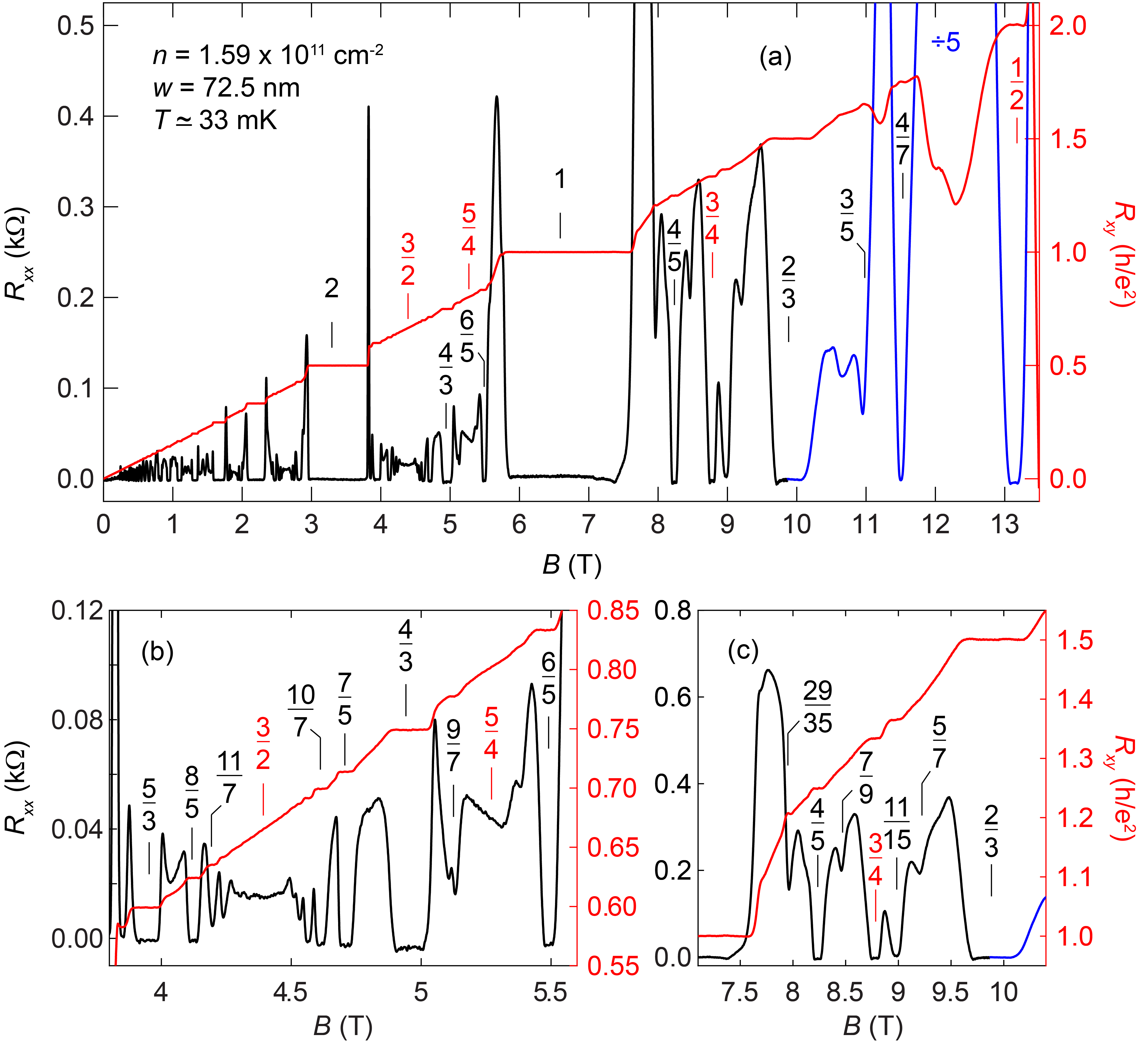}
\caption{Full-field ${R_{xx}}$ and ${R_{xy}}$ traces at ${n = 1.59}\times10^{11}\ \rm{cm}^{-2}$. Expanded views of $R_{xx}$ and $R_{xy}$ around (b) $\nu = 3/2$ and 5/4, and (c) $\nu = 3/4$. The 2DES still shows 1C behavior around $\nu = 3/2$, but becomes 2C at smaller fillings $(\nu<1)$ based on the weakening of \textit{odd-numerator} FQHSs at $\nu = 7/9$, 5/7 and 3/5, and the emergence of 2C FQHSsat $\nu = 3/4$, 11/15 and 29/35. Even-numerator FQHSs at $\nu = 2/3$, 4/5 and 4/7 still persist. A $\nu = 1/2$ FQHS is also observed, presumably in the same universality class as the 2C, $\Psi_{331}$ state \cite{Singh.PRL.2025}.}
\end{figure*}

\begin{figure*}[t!]
\hypertarget{app1-3}{}
\includegraphics[width=1\columnwidth]{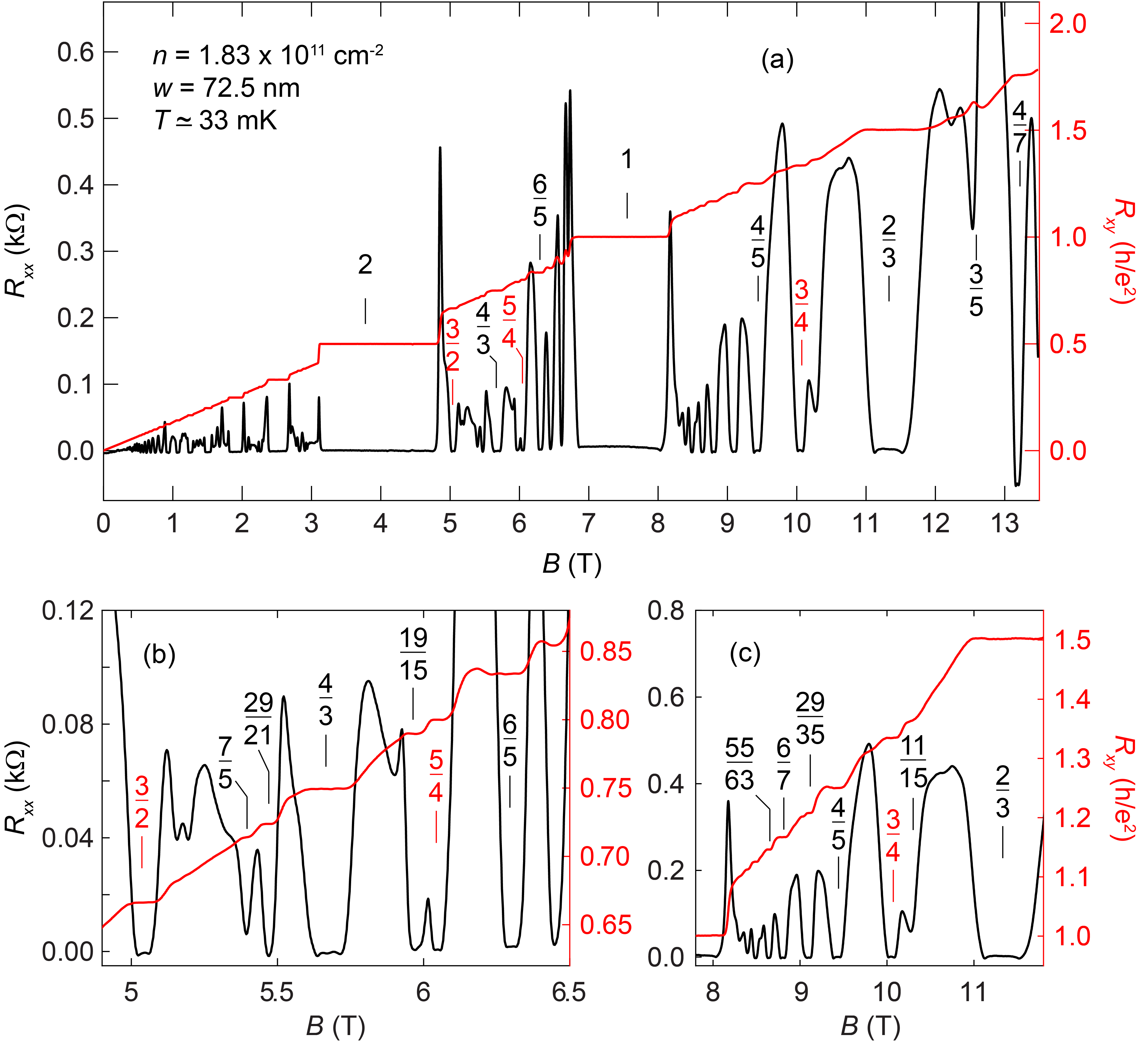}
\caption{Full-field ${R_{xx}}$ and ${R_{xy}}$ traces at ${n = 1.83}\times10^{11}\ \rm{cm}^{-2}$. Expanded views of $R_{xx}$ and $R_{xy}$ around (b) $\nu = 3/2$ and 5/4, and (c) $\nu = 3/4$. The 2DES is 2C for $\nu \leq 2$ as evinced from the emergence of balanced and imbalanced bilayer FQHSs.}
\end{figure*}

\begin{figure*}[b!]
\hypertarget{pd_n}{}
\includegraphics[width=1\columnwidth]{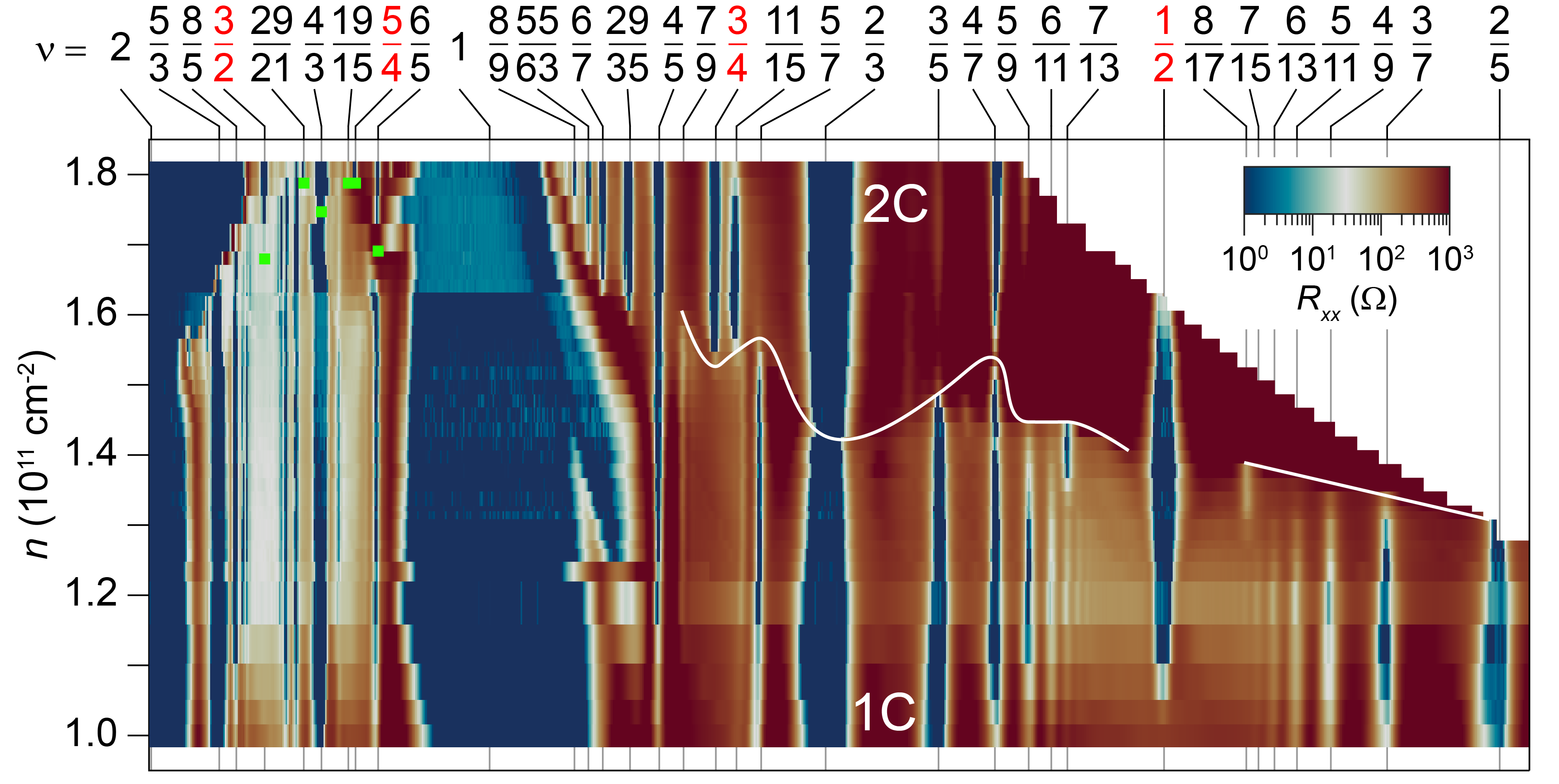}
\caption{Color-scale phase diagram of the 2DES with the color scale representing $R_{xx}$ as a function of \textit{n} (y-axis) and $1/\nu$ (x-axis). Note that the x-axis ticks mark the expected positions of the indicated FQHSs. Several FQHS transitions at odd-denominator $\nu$ can be observed, most notably at $\nu = 4/7$, 3/5, 5/7 and 7/9. The white curve is a guide-to-the-eye to highlight these transitions. Also, at high densities, insulating phases emerge at low fillings and engulf the FQHSs, including the $\nu = 1/2$ FQHS at the highest densities.}
\end{figure*}

\begin{figure*}[t!]
\hypertarget{pd_alpha}{}
\includegraphics[width=1\columnwidth]{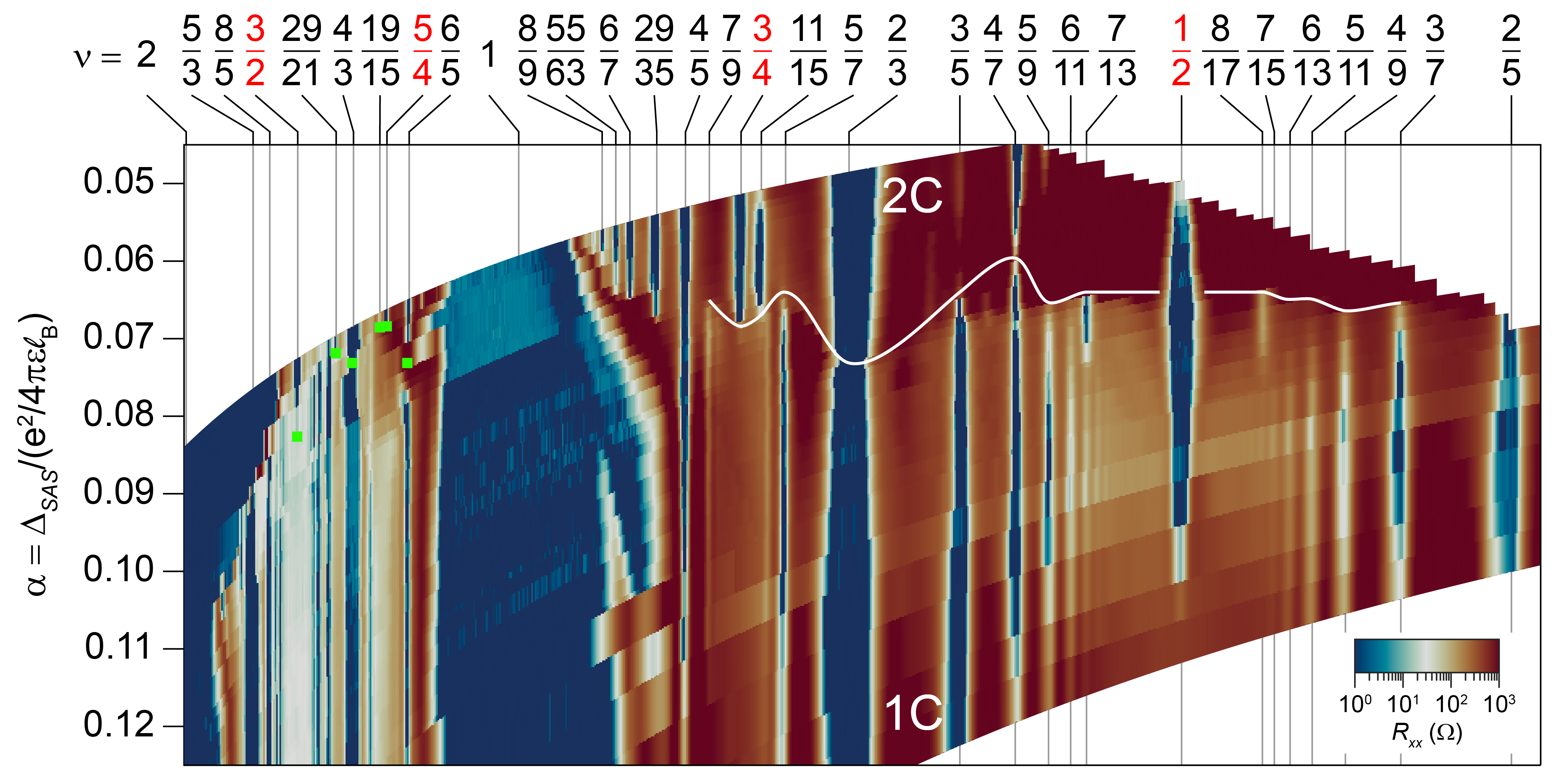}
\caption{Data of Fig. \protect\hyperlink{pd_n}{S4} are replotted as a function of the parameter $\alpha$. The white curve is presented as a guide-to-the-eye that follows the 1C to 2C transitions of the FQHSs for $\nu \leq 1$, which all occur around $\alpha \simeq 0.064$. Green squares are used to mark the emergence of the $\nu = 3/2$, 29/21, 19/15 and 5/4 FQHSs, and the 1C to 2C transitions of the even-numerator 4/3 and 6/5 FQHSs, and as $\alpha$ is decreased by increasing the density.}
\end{figure*}

\begin{figure*}[t!]
\hypertarget{doverlb}{}
\includegraphics[]{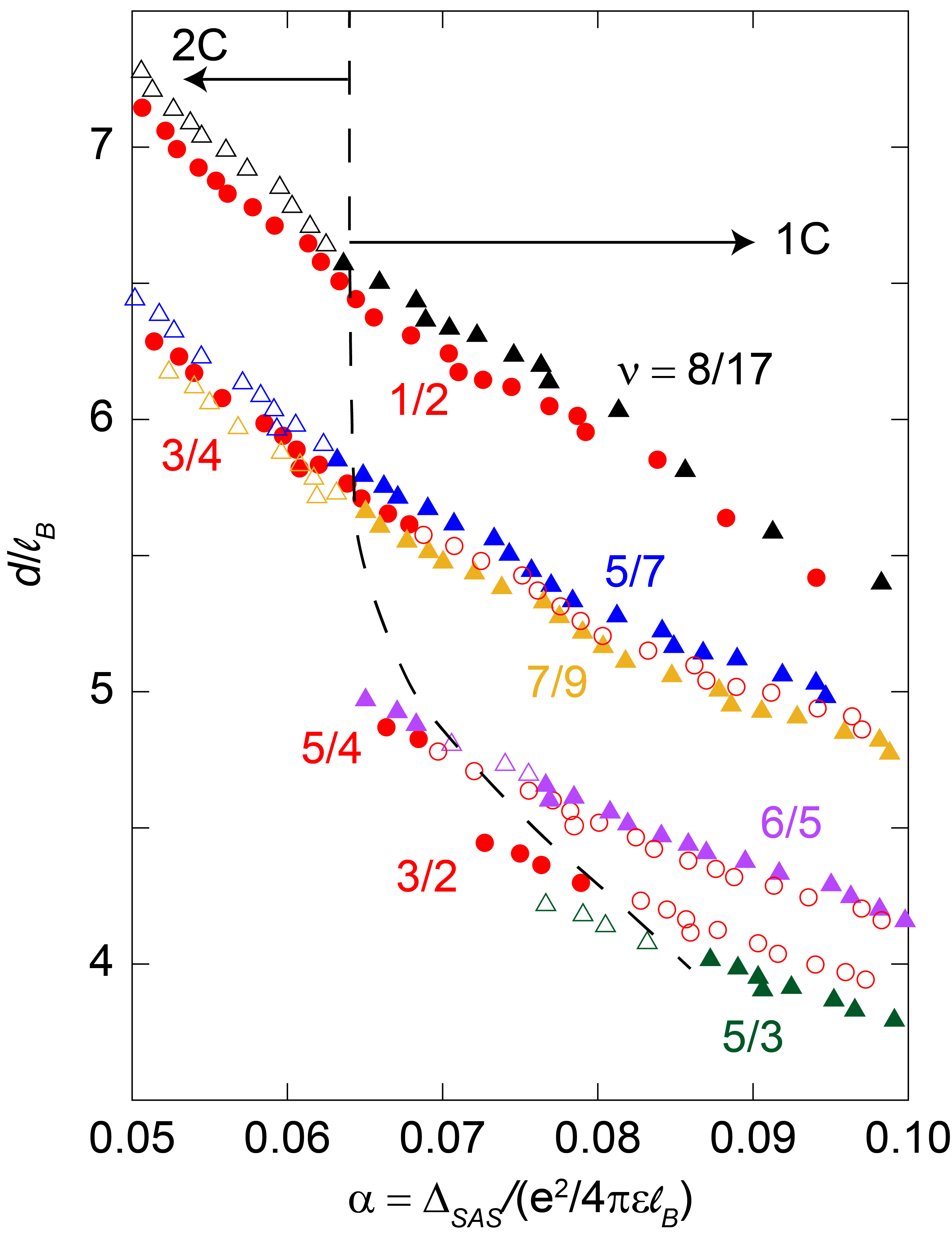}
\caption{Trajectories of the ground states at $\nu = 1/2$, 3/4, 5/4 and 3/2 in the $d/\ell_B$ vs $\alpha$ phase space. Also shown are the trajectories of nearby odd-denominator states. The even-numerator FQHS at $\nu = 6/5$ makes a 1C to 2C transition as $\alpha$ is reduced. Solid symbols denote the presence of a FQHS, while open symbols represent its absence. The 1C to 2C transitions for $\nu<1$ occur around $\alpha \simeq 0.064$. For $\nu > 1$, we observe that the 1C to 2C transitions move to larger $\alpha$ and smaller $d/\ell_B$. The dashed black line is a guide-to-the-eye which roughly follows the 1C - 2C transition of the 2DES, and converges to $\alpha \simeq 0.064$ for $\nu<1$.}
\end{figure*}

\begin{figure*}[h!]
\hypertarget{fivequarterevolution}{}
\includegraphics[width = 0.6\columnwidth]{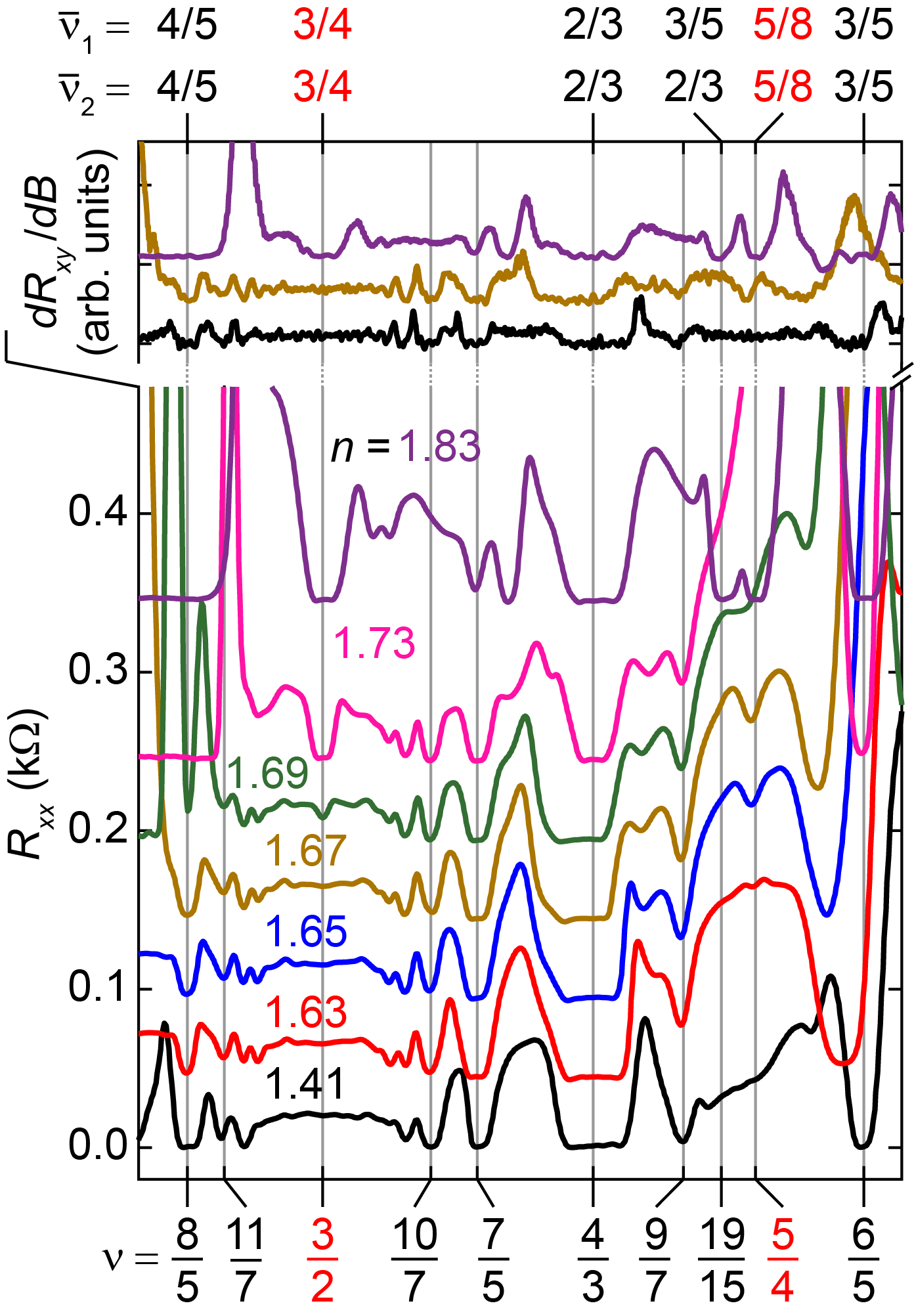}
\caption{$R_{xx}$ traces depicting the evolution around $\nu = 3/2$ and 5/4 shown over an expanded y-axis scale. Also shown at the top are the derivative of $R_{xy}$ color coded to the density. Deviation of $R_{xy}$ from the classical Hall value, accompanied with a minimum in $R_{xx}$ for $n = 1.67$, suggest a developing FQHS at $\nu = 5/4$.}
\end{figure*} 

\begin{figure*}[h!]
\hypertarget{fivequarter1C}{}
\includegraphics[width = 0.6\columnwidth]{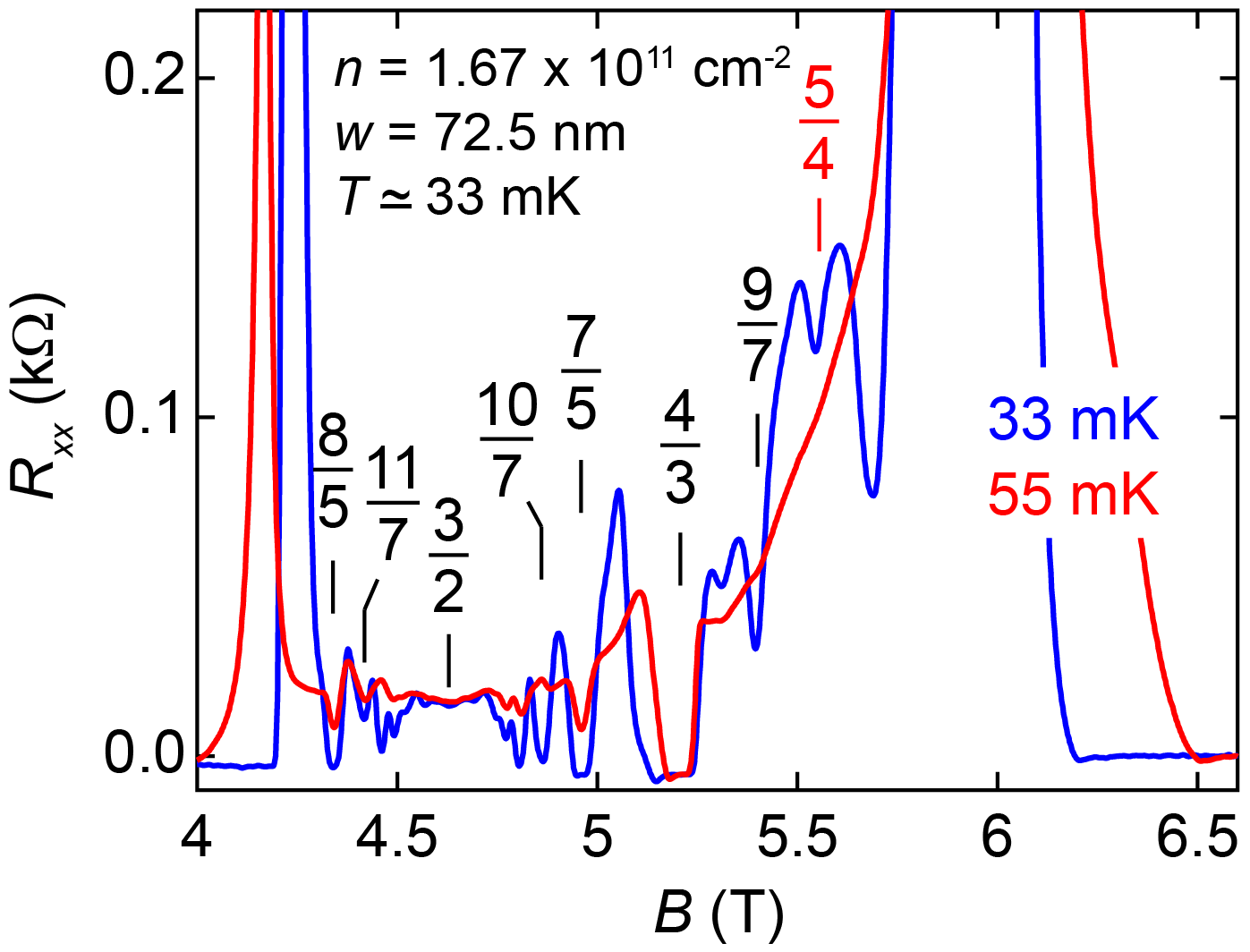}
\caption{Temperature dependence of the $\nu = 5/4$ FQHS observed at $n = 1.67$. This state is observed only at the lowest temperature ($\simeq 33$ mK), and disappears quickly as the temperature is raised to 55 mK. The temperature dependence suggests that the feature has a correlated origin.}
\end{figure*}